\documentclass[12pt,a4paper]{article}
\usepackage{graphics,graphicx}
\usepackage{amssymb,epsfig,amsmath,euscript,array}
\usepackage{cite}

\makeatletter
\@addtoreset{equation}{section}
\makeatother

\newcounter{multieqs}

\newcommand{\be}{\begin{equation}}
\newcommand{\ee}{\end{equation}}

\newcommand{\bm}[1]{\mbox{\boldmath $#1$}}

\newcommand{\kslash}{k \!\!\! / }

\newcommand{\lslash}{l \!\! / }
\newcommand{\Pslash}{P \!\!\!\! / }

\newcommand{\islash}{i \!\!\! / }
\newcommand{\jslash}{j \!\!\! / }
\newcommand{\aslash}{a \!\!\! / }
\newcommand{\bslash}{{b \hspace{-6pt} \slash} }

\newcommand{\onslash}{1 \!\!\! / }
\newcommand{\twslash}{2 \!\!\!/ }
\newcommand{\thslash}{3 \!\!\!/ }
\newcommand{\foslash}{4 \!\!\! / }
\newcommand{\fislash}{5 \!\!\! / }

\newcommand{\mslash}{m \!\!\! / }

\def\bd{\begin{document}}
\def\ed{\end{document}}
\def\nn{\nonumber}
\def\bea{\begin{eqnarray}}
\def\eea{\end{eqnarray}}
\def\eps{\epsilon}

\def\ab{(ijab)}
\def\ba{(ijba)}
\def\ijab{{\tr}_{+}(\islash\, \jslash\, \aslash \, \bslash)}
\def\ijba{{\tr}_{+}(\islash\, \jslash\, \bslash \, \aslash)}
\def\ijaP{{\tr}_{+}(\islash\, \jslash\, \aslash \, \Pslash)}
\def\ijPLa{{\tr}_{+}(\islash\, \jslash\, \Pslash_L \, \aslash)}
\def\ijaPL{{\tr}_{+}(\islash\, \jslash\, \aslash \, \Pslash_L)}
\def\ijPLza{{\tr}_{+}(\islash\, \jslash\, \Pslash_{L;z} \, \aslash)}
\def\ijaPLz{{\tr}_{+}(\islash\, \jslash\, \aslash \, \Pslash_{L;z})}
\def\ijPa{{\tr}_{+}(\islash\, \jslash\, \Pslash \, \aslash)}
\def\iaPb{{\tr}_{+}(\islash\, \aslash\, \Pslash \, \bslash)}
\def\ibPa{{\tr}_{+}(\islash\, \bslash\, \Pslash \, \aslash)}
\def\ijPmu{{\tr}_{+}(\islash\, \jslash\, \Pslash \, \mu)}
\def\ibmuP{{\tr}_{+}(\islash\, \bslash\, \mu \, \Pslash)}
\def\ibmua{{\tr}_{+}(\islash\, \bslash\, \mu \, \aslash)}
\def\iamub{{\tr}_{+}(\islash\, \aslash\, \mu \, \bslash)}
\def\jaPb{{\tr}_{+}(\jslash\, \aslash\, \Pslash \, \bslash)}
\def\ijmuP{{\tr}_{+}(\islash\, \jslash\, \mu \, \Pslash)}
\def\ijmum{{\tr}_{+}(\islash\, \jslash\, \mu \, \mslash)}
\def\ijmmu{{\tr}_{+}(\islash\, \jslash\, \mslash \, \mu)}
\def\ijmP{{\tr}_{+}(\islash\, \jslash\, \mslash \, \Pslash)}
\def\iabP{{\tr}_{+}(\islash\, \aslash\, \bslash \, \Pslash)}
\def\ijbP{{\tr}_{+}(\islash\, \jslash\, \bslash \, \Pslash)}
\def\jbPa{{\tr}_{+}(\jslash\, \bslash\, \Pslash \, \aslash)}
\def\ijPb{{\tr}_{+}(\islash\, \jslash\, \Pslash \, \bslash)}
\def\jbmua{{\tr}_{+}(\jslash\, \bslash\, \mu \, \aslash)}

\def\loablt{ {\tr}_{+}(\lslash_1\, \aslash \, \bslash\, \lslash_2)}

\def\ijlolt{{\tr}_{+}(\islash\, \jslash\, \lslash_1 \, \lslash_2)}
\def\ijltlo{{\tr}_{+}(\islash\, \jslash\, \lslash_2 \, \lslash_1)}
\def\ibloa{{\tr}_{+}(\islash\, \bslash\, \lslash_1 \, \aslash)}
\def\jaltb{{\tr}_{+}(\jslash\, \aslash\, \lslash_2 \, \bslash)}
\def\ialtb{{\tr}_{+}(\islash\, \aslash\, \lslash_2 \, \bslash)}
\def\bltloa{{\tr}_{+}(\bslash\, \lslash_2\, \lslash_1 \, \aslash)}
\def\jbloa{{\tr}_{+}(\jslash\, \bslash\, \lslash_1 \, \aslash)}
\def\ibPb{{\tr}_{+}(\islash\, \bslash\, \Pslash \, \bslash)}
\def\ijltb{{\tr}_{+}(\islash\, \jslash\, \lslash_2 \, \bslash)}

\def\ijloa{{\tr}_{+}(\islash\, \jslash\,  \lslash_1 \, \aslash)}
\def\ijblt{{\tr}_{+}(\islash\, \jslash\,  \bslash \, \lslash_2)}

\def\jakb{{\tr}_{+}(\jslash\, \aslash\, \kslash \, \bslash)}
\def\iakb{{\tr}_{+}(\islash\, \aslash\, \kslash \, \bslash)}

\def\tofo{{\tr}_{+}(\onslash\, \thslash\, \twslash \, \foslash)}
\def\foto{{\tr}_{+}(\onslash\, \thslash\, \foslash \, \twslash)}
\def\tofi{{\tr}_{+}(\onslash\, \thslash\, \twslash \, \fislash)}
\def\fito{{\tr}_{+}(\onslash\, \thslash\, \fislash \, \twslash)}

\def\lrangle#1#2{\langle #1\,#2\rangle}

\def\Li{{$\rm Li}_2$}

\let\bm=\bibitem
\let\la=\label

\def\npb#1#2#3{Nucl. Phys. {\bf{B#1}} #3 (#2)}
\def\plb#1#2#3{Phys. Lett. {\bf{#1B}} #3 (#2)}
\def\prl#1#2#3{Phys. Rev. Lett. {\bf{#1}} #3 (#2)}
\def\prd#1#2#3{Phys. Rev. {D \bf{#1}} #3 (#2)}
\def\cmp#1#2#3{Comm. Math. Phys. {\bf{#1}} #3 (#2)}
\def\cqg#1#2#3{Class. Quantum Grav. {\bf{#1}} #3 (#2)}
\def\nppsa#1#2#3{Nucl. Phys. B (Proc. Suppl.) {\bf{#1A}}#3 (#2)}
\def\ap#1#2#3{Ann. of Phys. {\bf{#1}} #3 (#2)}
\def\ijmp#1#2#3{Int. J. Mod. Phys. {\bf{A#1}} #3 (#2)}
\def\rmp#1#2#3{Rev. Mod. Phys. {\bf{#1}} #3 (#2)}
\def\mpla#1#2#3{Mod. Phys. Lett. {\bf A#1} #3 (#2)}
\def\jhep#1#2#3{J. High Energy Phys. {\bf #1} #3 (#2)}
\def\atmp#1#2#3{Adv. Theor. Math. Phys. {\bf #1} #3 (#2)}
\newcommand{\EQ}[1]{\begin{equation} #1 \end{equation}}
\newcommand{\AL}[1]{\begin{subequations}\begin{align} #1 \end{align}\end{subequations}}
\newcommand{\SP}[1]{\begin{equation}\begin{split} #1 \end{split}\end{equation}}
\newcommand{\ALAT}[2]{\begin{subequations}\begin{alignat}{#1} #2 \end{alignat}
                        \end{subequations}}
\def\beqa{\begin{eqnarray}}
\def\eeqa{\end{eqnarray}}
\def\beq{\begin{equation}}
\def\eeq{\end{equation}}
\def\sst{\scriptscriptstyle}
\def\thetabar{\bar\theta}
\def\Tr{{\rm Tr}}
\def\one{\mbox{1 \kern-.59em {\rm l}}}
 \def\Nh{\hat{N}}

\def\a{\alpha}      \def\da{{\dot\alpha}}
\def\b{\beta}       \def\db{{\dot\beta}}
\def\g{\gamma}  \def\G{\Gamma}  \def\cdt{\dot\gamma}
\def\d{\delta}  \def\D{\Delta}  \def\ddt{\dot\delta}
\def\e{\epsilon}        \def\vare{\varepsilon}
\def\f{\phi}    \def\F{\Phi}    \def\vvf{\f}
\def\h{\eta}
\def\k{\kappa}
\def\l{\lambda} \def\L{\Lambda}
\def\m{\mu} \def\n{\nu}
\def\o{\omega}
\def\p{\pi} \def\P{\Pi}
\def\r{\rho}
\def\s{\sigma}  \def\S{\Sigma}
\def\t{\tau}
\def\th{\theta} \def\Th{\Theta} \def\vth{\vartheta}
\def\X{\Xeta}
\def\z{\zeta}

\def\cA{{\cal A}} \def\cB{{\cal B}} \def\cC{{\cal C}}
\def\cD{{\cal D}} \def\cE{{\cal E}} \def\cF{{\cal F}}
\def\cG{{\cal G}} \def\cH{{\cal H}} \def\cI{{\cal I}}
\def\cJ{{\cal J}} \def\cK{{\cal K}} \def\cL{{\cal L}}
\def\cM{{\cal M}} \def\cN{{\cal N}} \def\cO{{\cal O}}
\def\cP{{\cal P}} \def\cQ{{\cal Q}} \def\cR{{\cal R}}
\def\cS{{\cal S}} \def\cT{{\cal T}} \def\cU{{\cal U}}
\def\cV{{\cal V}} \def\cW{{\cal W}} \def\cX{{\cal X}}
\def\cY{{\cal Y}} \def\cZ{{\cal Z}}

\def\ua{\underline{\alpha}}
\def\ub{\underline{\phantom{\alpha}}\!\!\!\beta}
\def\uc{\underline{\phantom{\alpha}}\!\!\!\gamma}
\def\um{\underline{\mu}}
\def\ud{\underline\delta}
\def\ue{\underline\epsilon}
\def\una{\underline a}\def\unA{\underline A}
\def\unb{\underline b}\def\unB{\underline B}
\def\unc{\underline c}\def\unC{\underline C}
\def\und{\underline d}\def\unD{\underline D}
\def\une{\underline e}\def\unE{\underline E}
\def\unf{\underline{\phantom{e}}\!\!\!\! f}\def\unF{\underline F}
\def\unm{\underline m}\def\unM{\underline M}
\def\unn{\underline n}\def\unN{\underline N}
\def\unp{\underline{\phantom{a}}\!\!\! p}\def\unP{\underline P}
\def\unq{\underline{\phantom{a}}\!\!\! q}
\def\unQ{\underline{\phantom{A}}\!\!\!\! Q}
\def\unH{\underline{H}}

\def\As {{A \hspace{-6.4pt} \slash}\;}
\def\bs {{b \hspace{-6.4pt} \slash}\;}
\def\Ds {{D \hspace{-6.4pt} \slash}\;}
\def\ds {{\del \hspace{-6.4pt} \slash}\;}
\def\ss {{\s \hspace{-6.4pt} \slash}\;}
\def\ks {{ k \hspace{-6.4pt} \slash}\;}
\def\ps {{p \hspace{-6.4pt} \slash}\;}
\def\pas {{{p_1} \hspace{-6.4pt} \slash}\;}
\def\pbs {{{p_2} \hspace{-6.4pt} \slash}\;}
\def\Ps {{P \hspace{-6.4pt} \slash}\;}
\def\Qs {{Q \hspace{-6.4pt} \slash}\;}

\def\Fh{\hat{F}}
\def\Vh{\hat{V}}
\def\Xh{\hat{X}}
\def\ah{\hat{a}}
\def\xh{\hat{x}}
\def\yh{\hat{y}}
\def\ph{\hat{p}}
\def\xih{\hat{\xi}}
\def\psit{\tilde{\psi}}
\def\Psit{\tilde{\Psi}}
\def\tht{\tilde{\th}}
\def\lt{\tilde{\lambda}}
\def\llt{\tilde{l}}
\def\At{\tilde{A}}
\def\Qt{\tilde{Q}}
\def\Rt{\tilde{R}}
\def\Nt{\tilde{N}}

\def\at{\tilde{a}}
\def\st{\tilde{s}}
\def\ft{\tilde{f}}
\def\pt{\tilde{p}}
\def\qt{\tilde{q}}
\def\vt{\tilde{v}}
\def\nt{\tilde{n}}

\def\delb{\bar{\partial}}
\def\bz{\bar{z}}
\def\bD{\bar{D}}
\def\bB{\bar{B}}

\def\bk{{\bf k}}
\def\bl{{\bf l}}
\def\bp{{\bf p}}
\def\bq{{\bf q}}
\def\br{{\bf r}}
\def\bx{{\bf x}}
\def\by{{\bf y}}
\def\bR{{\bf R}}
\def\bV{{\bf V}}

\def\d{\delta}\def\D{\Delta}\def\ddt{\dot\delta}
\def\pa{\partial} \def\del{\partial}
\def\xx{\times}
\def\uno{\mbox{1 \kern-.59em {\rm l}}}
\def\trp{^{\top}}
\def\inv{^{-1}}
\def\dag{{^{\dagger}}}
\def\pr{^{\prime}}
\def\lan{\langle}
\def\ran{\rangle}
\def\rar{\rightarrow}
\def\lar{\leftarrow}
\def\lrar{\leftrightarrow}
\newcommand{\0}{\,\!}      
\def\one{1\!\!1\,\,}
\def\im{\imath}
\def\jm{\jmath}
\newcommand{\tr}{\mbox{tr}}
\newcommand{\slsh}[1]{/ \!\!\!\! #1}
\def\vac{|0\rangle}
\def\lvac{\langle 0|}
\def\hlf{\frac{1}{2}}
\def\ove#1{\frac{1}{#1}}
\def\Box{\square}
\def\ZZ{\mathbb{Z}}
\def\CC#1{({\bf #1})}
\def\bcomment#1{}
\def\bfhat#1{{\bf \hat{#1}}}
\def\VEV#1{\left\langle #1\right\rangle}
\newcommand{\ex}[1]{{\rm e}^{#1}} \def\ii{{\rm i}}
\def\rr{{\rm r}} \def\rs{{\rm s}}\def\rv{{\rm v}}
\def\ri{{\rm i}}\def\rj{{\rm j}}
\newcommand{\lrbrk}[1]{\left(#1\right)}
\newcommand{\sfrac}[2]{{\textstyle\frac{#1}{#2}}}

\def\Li{{\rm Li}_2}

\font\mybb=msbm10 at 12pt
\def\bb#1{\hbox{\mybb#1}}

\font\myBB=msbm10 at 18pt
\def\BB#1{\hbox{\myBB#1}}

\begin{document}

\begin{flushright}
hep-th/0701187 \\
QMUL-PH-07-02
\end{flushright}

\vspace{20pt}

\begin{center}

{\Large \bf Recursion Relations for}
\\
\vspace{0.3cm} {\Large \bf One-Loop Gravity Amplitudes} \vspace{12pt}
\vspace{33pt}

{\bf {\mbox{Andreas Brandhuber, Simon McNamara, Bill Spence and Gabriele Travaglini}}}%
\footnote{{\sffamily \{\tt a.brandhuber, s.mcnamara, w.j.spence, g.travaglini\}@qmul.ac.uk }}

{\em Centre for Research in String Theory\\
Department of Physics\\
Queen Mary, University of
London\\
Mile End Road, London, E1 4NS\\
United Kingdom}
\vspace{40pt}

{\bf Abstract}

\end{center}

\noindent

We study the application of recursion relations to the calculation of finite one-loop gravity
amplitudes. It is shown explicitly that the known  five, and six graviton one-loop
amplitudes for which the external legs have identical outgoing helicities, and the four graviton
amplitude with helicities $-+++$
can be derived from simple recursion relations. The latter amplitude is
derived by introducing a one-loop three-point vertex of gravitons of positive helicity, which
is the counterpart in gravity of the one-loop three-plus vertex in Yang-Mills.
We show  that new issues arise for the five point amplitude with helicities $-++++$, where the application
of known methods does not appear to work, and we discuss possible resolutions.

\vspace{0.5cm}

\setcounter{page}{0}
\thispagestyle{empty}

\newpage
\section{Introduction}
\setcounter{footnote}{0}

The twistor string proposal of Witten \cite{Witten:2003nn}
has inspired many new techniques for the calculation of
scattering amplitudes in gauge theory and gravity (see the reviews
\cite{Cachazo:2005ga, Brandhuber:2006vh}, and references therein).
Amongst these new developments, an important conceptual and practical advance
was the derivation of recursion relations in gauge theories. This was achieved in
\cite{Britto:2004ap,  Britto:2005fq},
incorporating insights from \cite{BDDK, BCF, BDK, RSV}.
These BCFW recursion relations proved to be
a very efficient technique for calculating scattering amplitudes, and new
results at tree level in gauge theory were rapidly found
\cite{Luo:2005rx, Luo:2005my, Badger:2005zh}.
The application to loop level amplitudes proved to be more involved however.
A notable step was taken in \cite{Bern:2005hs}, where it was shown that
the introduction of a new one-loop three vertex made possible
a derivation of one-loop amplitudes using recursive techniques.
However, for ``non-standard" cases
it was shown that correction terms to na\"{i}ve recursive rules were
needed for the process to work; these cases occur when the amplitudes develop
single poles masked by double poles as momenta are continued to complex values.
It was not immediately clear how this approach could be systematised in full generality,
but further work at one-loop level clarified a number of issues and made further progress
\cite{Bern:2005ji, Bern:2005cq, Berger:2006ci, Berger:2006vq, Bern:2005hh}.

For the case of gravity amplitudes, tree-level recursion relations have also been found
\cite{Bedford:2005yy, Cachazo:2005ca}.
This involved the discovery of some new tree amplitudes and new forms of known
tree amplitudes. Again it was clear that these relations were of considerable
practical use, leading to much simpler derivations, as well as final forms,
of amplitudes.
It is natural to ask if quantum gravity amplitudes can also be studied using recursion
relations. This might be relevant given recent interesting results and conjectures
concerning $\cN=8$ supergravity (see  \cite{Bjerrum-Bohr:2006yw, Green:2006gt, Bern:2006kd,Green:2006yu} and references therein).
In this letter we study this
question, showing to what extent the new quantum gauge theory recursion
techniques can be applied to gravity, and indicating where this
approach breaks down and why.

As usual we write
amplitudes in the spinor-helicity formalism and analytically continue
the spinors to complex momenta where needed in the arguments.
The BCFW recursion relations are based on two very general
properties of amplitudes -- analyticity
and factorisation on multi-particle poles.
One considers the following deformation of an amplitude which
shifts the spinors of two of the $n$ massless external particles,
labelled $i$ and $j$, and involves a complex parameter $z$,
\begin{eqnarray}\label{eq:bcfwshifts}
\lambda_i&\to&\lambda_i
\ ,
\nonumber\\
\widetilde{\lambda}_i&\to&\widetilde{\lambda}_i-z\widetilde{\lambda}_j
\ ,
\nonumber\\
\lambda_j&\to&\lambda_j+z\lambda_i
\ ,
\nonumber\\
\widetilde{\lambda}_j&\to&\widetilde{\lambda}_j
\ .
\end{eqnarray}
This deformation does not make sense for real momenta in
Minkowski space which satisfy
$\widetilde{\lambda}=\pm\bar{\lambda}$,
but is perfectly consistent for complex kinematics.
Under the shifts \eqref{eq:bcfwshifts}, the corresponding
momenta $p_i(z)$ and $p_j(z)$ remain on-shell for all complex $z$,
and $p_i(z)+p_j(z)=p_i(0)+p_j(0)$.
Hence the quantity $A(1,\ldots,p_i(z),\ldots,p_j(z),\ldots,n)$
is a well-defined on-shell amplitude for all $z$.%
\footnote{One can consider more general deformations than
(\ref{eq:bcfwshifts}) of course --
for example more exotic shifts have shown  \cite{Risager:2005vk}
that the
tree-level MHV rules  \cite{csw}
are an instance of BCFW recursion, and multiple shifts
have been used to eliminate boundary terms in the generalisation
of BCFW recursion to one-loop QCD amplitudes \cite{Bern:2005hs}.}

The analytic structure of the $z$-dependent amplitude $A(z)$ is then used to
calculate the physical amplitude  $A(0)$.
Specifically, the recursion relation can be derived from considering the
following contour integral, where the contour $C$ is the circle at infinity:
\begin{equation}
\label{van}
\frac{1}{2\pi i} \oint_{C} dz \frac{A(z)}{z}
\ .
\end{equation}
The integral in \eqref{van} vanishes if we assume
that $A(z) \to 0$ as $z \to \infty$. It  then follows from Cauchy's residue theorem
that we can write the amplitude we wish to calculate, $A(0)$,  as a sum of residues
of $A(z)/z$:
\begin{equation}
A(0)\ =\ -\sum_{\substack{\textrm{poles of $A(z)/z$}\\\textrm{excluding $z\!=\!0$}}}
\textrm{Res} \left\{\frac{A(z)}{z}\right\}
\ .
\end{equation}
For tree-level Yang-Mills $A(z)$ has only simple poles.
A pole at $z\!=\!z_{P}$ is associated with a shifted momentum
$P(z) := P + z \eta$ becoming null.
The residue at this pole is then obtained by factorising the shifted amplitude on
this pole,
\begin{equation}
\textrm{Res} \left\{\frac{A(z)}{z}\right\}\, =
\,
\sum_h A_L^h(z_{P})
 \frac{i}{P^2} A_R^{-h}(z_{P})
 \ ,
\end{equation}
where the sum is over the possible  assignments of the helicity $h$ of the intermediate state.
The left and right shifted amplitudes $A_L$ and $A_R$ are, of course,
only defined for $z\!=\!z_{P}$ when
$P(z)$ is null.
The intermediate propagator is evaluated
with unshifted kinematics.
Since a momentum invariant involving both (or neither) of the shifted legs $i$
and $j$ does not give rise to a pole in $z$, the shifted legs $i$ and
$j$ must always appear on opposite sides of the factorisation.

Now consider the generalisation of the BCFW recursion relations which was used to derive
the rational parts of one-loop gluon amplitudes in QCD \cite{Bern:2005hs}.
Whilst the structure of multi-particle factorisation of tree-level amplitudes implies that in
general only simple poles result from performing shifts on a tree-level amplitude,
this is not the case for the rational terms of one-loop Yang-Mills amplitudes.
The reason for this is that in real momenta the one-loop splitting functions have only simple
poles, however in complex momenta the one-loop splitting functions with helicities
$+++$ and $---$ develop double poles.
Since Yang-Mills tree-level amplitudes with more than three legs and
less than two negative
helicity gluons vanish, the one-loop amplitude with all legs of positive helicity is finite and
has no multi-particle poles. Thus the shifted all-plus amplitude only has
simple poles coming from the collinear singularities of the tree-level $-++$ splitting amplitude.
Once shifts without a boundary term have been found, the all-plus amplitudes
can then be constructed recursively by sewing all-plus loop amplitudes with
fewer legs to three-gluon tree amplitudes \cite{Bern:2005hs}.

Remarkably, the ideas of BCFW recursions are also applicable to more general QCD
one-loop amplitudes such as the amplitude with a single negative-helicity gluon.
As can be seen by performing the BCFW shifts on a known amplitude,
there is an added complication in this one-loop recursion, as performing a shift
results in the appearance of a double pole. As explained in \cite{Bern:2005hs}
these double poles are associated with the appearance of three-point all-plus
one-loop vertices. Cauchy's residue theorem does, of course,
extend to this case --  although the double-pole in $A(z)$ does not have
a residue,  we are integrating $A(z)/z$ which does have a residue,
\begin{equation}
\begin{array}{c} \textrm{Res} \\ z=a \end{array} \left\{\frac{1}{z(z-a)^2}\right\}=-\frac{1}{a^2}.
\end{equation}
Factorisation at a double pole will therefore be schematically of the form
\begin{equation}
A_L \frac{1}{(P^2)^2} A_R
\ .
\end{equation}
It is clear even on dimensional grounds that $A_L$ and $A_R$
cannot both be amplitudes,
hence the factorisation on a double pole  will involve
a {\it vertex} with the dimensions of an amplitude times a momentum squared.
This may seem puzzling at first sight,
but it can be understood from the structure of the one-loop three-point vertex
used for obtaining one-loop splitting amplitudes
\begin{equation}
A_3^{(1)} (1^+,2^+,3^+) = -i\frac{N_p}{96 \pi^2} \frac{[12][23][31]}{K_{12}^2}
\ .
\end{equation}
This explicit formula shows that the three-point one-loop all-plus amplitude is either zero or infinite
even in complex momenta, as it involves both the holomorphic and anti-holomorphic
spinor variables. To compute the recursive double pole terms associated with the three-point
all-plus factorisations, Bern, Dixon and Kosower (BDK) proposed in  \cite{Bern:2005hs}
the use of the following vertex, which has the right
dimensions and is only a function of the $\widetilde{\lambda}$ variables:
\begin{equation}\label{eq:ymvertex}
V_3^{(1)} (1^+,2^+,3^+)=-\frac{i}{96 \pi^2} [12][23][31]
\ .
\end{equation}
In order to derive the single pole underneath  the double pole,
it was conjectured  that the single pole differs from the double pole by a
factor of the form
\begin{equation}
\label{eq:singleunderdouble}
S(a_1,\hat{K}^+,a_2) \, K^2 \, S(b_1,-\hat{K}^-,b_2)
\ ,
\end{equation}
where $K^2$ is the propagator responsible
for the pole in the shifted amplitude.
The soft functions in \eqref{eq:singleunderdouble}
are given by
\begin{eqnarray}
S(a,s^+,b)&=&\frac{\lan ab \ran}{\lan as \ran \lan sb \ran} \nonumber \\
S(a,s^-,b)&=&-\frac{[ab]}{[as] [sb]}\ .
\label{eq:softfactors}
\end{eqnarray}
The factor of $K^2$ in \eqref{eq:singleunderdouble}  cancels one of
the two $K^2$ factors in the double pole term, leaving a single pole.
The legs $b_1$ and $b_2$ are the external legs on the three-point
all-plus vertex.
Experimentation revealed that the legs $a_1$ and $a_2$ are to be
identified with the external legs of the tree amplitude part of
the recursive diagram which are colour adjacent to the propagator.

In the rest of the paper we will apply these ideas to study the rational
one-loop amplitudes in pure Einstein gravity.
In section 2 we consider the one-loop gravity amplitudes with all legs
of positive helicity for the five- and six-point
cases, showing that these can be correctly obtained with a suitable choice of shifts.
In section 3 we derive the four graviton amplitude with all but one
graviton of positive helicity.
We note that double poles appear in this case,
and show that the approach of \cite{Bern:2005hs} also works in the gravity case.
Specifically, we make use of a three-point one-loop vertex of positive helicity gravitons, which
generalises to the case of gravity the corresponding three-point amplitude in Yang-Mills theory.
The single pole underneath the double pole has a structure which is very similar to that found
for the Yang-Mills case. Perhaps surprisingly, we find that the Yang-Mills soft functions
 (rather than the gravity ones) are the objects in terms of which these single poles are expressed.
Finally in section 4 we turn to the five graviton amplitude with all but one graviton of positive helicity.
Using the techniques and rules applied successfully in the previous cases results in a formula for this
amplitude which turns out not to obey the essential requirements of symmetry, and collinear and soft
limit conditions. We discuss the possible sources of problems and potential resolutions.

\section{The all-plus amplitude}
An ansatz for the $n$ point one-loop amplitude in pure Einstein gravity in which all the
external gravitons have the same outgoing helicity was presented in \cite{Bern:1998xc}.
This agrees with explicit computations via $D$-dimensional unitarity cuts for
$n \leq 6$ \cite{Bern:1998sv}.
This amplitude corresponds to self-dual configurations of the field strength, and is
also related to the one-loop maximally helicity-violating (MHV) amplitude
in $\cN\!=\!8$ supergravity via the dimension-shifting relation of \cite{Bern:1998sv}.
It is
\begin{equation}\label{eq:allplus}
M_n^{(1)}(1^+,2^+,\ldots,n^+)=
-\frac{i}{(4 \pi)^2}\frac{1}{960} \sum_{\substack{1 \leq a < b \leq n \\ M, N}} h(a, M, b) h(b, N, a)
\tr[\kslash_a \kslash_M \kslash_b \kslash_N]^3
\ .
\end{equation}
In this formula $a$ and $b$ are massless legs and $M$ and $N$ are two sets forming a distinct
nontrivial partition of the remaining $n-2$ legs.
The first few half-soft functions $h(a, S, b)$ are given by
\begin{eqnarray}
h(a,\{1\},b)&=&\frac{1}{\lan a 1\ran^2 \lan 1 b\ran^2}\ , \nonumber\\
h(a,\{1,2\},b)&=&\frac{[12]}{\lan1 2\ran \lan a 1\ran \lan1 b\ran \lan a 2\ran \lan2 b\ran}\ , \nonumber\\
h(a,\{1,2,3\},b)&=&\frac{[12][23]}{\lan12\ran \lan23\ran \lan a1\ran \lan1b\ran \lan a3\ran \lan3b\ran}
+\frac{[23][31]}{\lan23\ran \lan31\ran \lan a2 \ran \lan2b\ran \lan a1\ran \lan1b\ran} \nonumber \\
&+&
\frac{[31][12]}{\lan31\ran \lan12\ran \lan a3\ran \lan3b\ran \lan a2\ran \lan2b\ran}\ .
\end{eqnarray}
Inspection of the all-plus amplitude  above shows that the standard
BCFW shifts \cite{Britto:2005fq} give a boundary term.
The following shifts, however, do not produce a boundary term:
\begin{eqnarray}\label{eq:risagershifts}
\hat{\lambda}_1&=&\lambda_1+z[23]\eta
\ ,  \nonumber\\
\hat{\lambda}_2&=&\lambda_2+z[31]\eta \ , \nonumber\\
\hat{\lambda}_3&=&\lambda_3+z[12]\eta
\ .
\end{eqnarray}
In the recursion relation we will write for
the five-point all-plus gravity amplitude we will make the convenient choice of
$\eta\!=\!\lambda_4+\lambda_5$
For the six-point all-plus amplitude recursion relation,
we will set  $\eta\!=\!\lambda_4+\lambda_5+\lambda_6$.
Applying the shifts (\ref{eq:risagershifts}) to the all-plus amplitude
(\ref{eq:allplus}) gives a shifted amplitude $M_n^{(1)}(z)$ with only simple poles.
In the following we show that the residues at these poles can be
computed from standard recursion relation diagrams.

\subsection{The five-point all-plus amplitude}

We will now use the on-shell one-loop recursion relation to re-derive the known five-point
all-plus amplitude from the four-point all-plus amplitude.
In this case, \eqref{eq:allplus} becomes
\begin{equation}\label{eq:4ptallplus}
M_4^{(1)}(1^+,2^+,3^+,4^+)=-\frac{i}{(4 \pi)^2}\frac{1}{60}
\left(\frac{[12] \, [34]}{\lan12\ran \lan34\ran}\right)^2
(s^2+st+t^2)\ ,
\end{equation}
where $s=(p_1+p_2)^2$ and $t=(p_2+p_3)^2$.
The amplitude in \eqref{eq:4ptallplus}
was first computed using string-based methods in \cite{Dunbar:1994bn}.

In the construction of the five-point all-plus amplitude, the shifts (\ref{eq:risagershifts})
give rise to nine different diagrams corresponding
to the nine different angle brackets that the shifts can make singular.
Our symmetric choice of $\eta = \lambda_4 +\lambda_5$ has the advantage that
there are only two distinct types of diagrams to compute,  the remaining ones  being
straightforward permutations of these two diagrams.
All the recursive diagrams contain a four-point one-loop amplitude joined to a tree-level
$++-$ amplitude.

\begin{figure}[ht]
\begin{center}
\scalebox{0.65}{\includegraphics{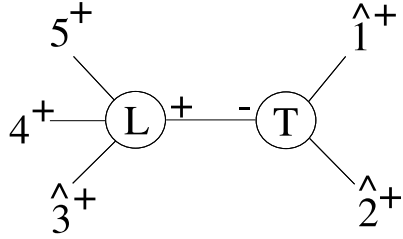}}
\end{center}
\caption{\it
The diagram in the recursive expression for $M_5^{(1)}(1^+,2^+,3^+,4^+,5^+)$ associated with the pole $\lan\hat{1}\hat{2}\ran=0$. The amplitude labelled by a T is a tree-level amplitude and the one labelled by L is a one-loop amplitude.}\label{figure-5ptallplus1}
\end{figure}

The first type of diagram has two shifted external legs attached to the tree-level
$++-$ diagram. There are three of these diagrams, corresponding to the simple-poles
$\lan\hat1 \hat2\ran\!=\!\lan\hat2 \hat3\ran\!=\!\lan\hat3 \hat1\ran\!=\!0$
in the shifted amplitude.
The diagram associated with the pole $\lan\hat1 \hat2\ran\!=\!0$ is drawn  in
Figure \ref{figure-5ptallplus1}.

\begin{figure}[ht]
\begin{center}
\scalebox{0.65}{\includegraphics{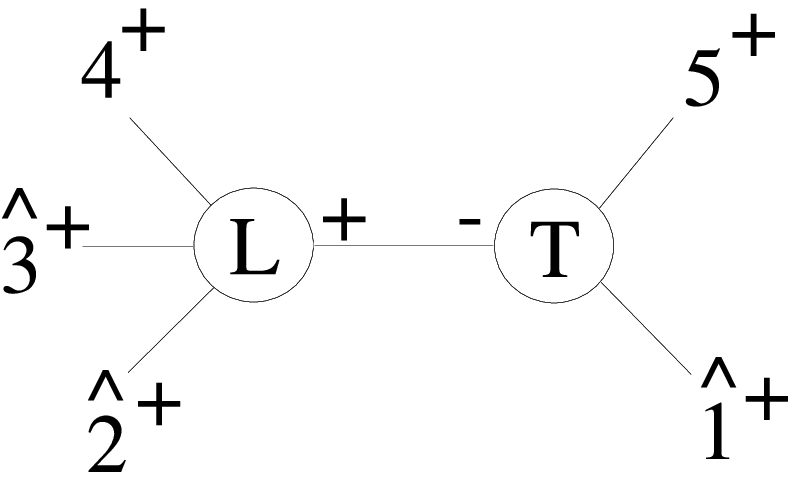}}
\end{center}
\caption{\it
The diagram in the recursive expression for $M_5^{(1)}(1^+,2^+,3^+,4^+,5^+)$ associated with the pole $\lan\hat{1}5\ran=0$.}
\label{figure-5ptallplus2}
\end{figure}

The second type of diagram has a shifted and an unshifted leg attached to the
tree-level $++-$ amplitude.
There are six such  diagrams, corresponding to  the simple-poles
$\lan\hat1 4\ran\!=\!\lan\hat1 5\ran\!=\!\lan\hat2 4\ran\!=
\!\lan\hat2 5\ran\!=\!\lan\hat3 4\ran\!=\!\lan\hat3 5\ran\!=\!0$
in the shifted amplitude.
We draw  in Figure \ref{figure-5ptallplus2} the diagram associated with the pole $\lan\hat1 5\ran\!=\!0$.

We start by considering the diagrams of the first type.
Specifically, the diagram in Figure \ref{figure-5ptallplus1} contributes
\begin{equation}\label{eq:5ptallplus1}
M_4^{(1)}(\hat{3}^+,4^+,5^+,\hat{K}_{12}^+) \frac{i}{K_{12}} M_3^{(0)}(\hat{1}^+,\hat{2}^+,-\hat{K}_{12}^-)\ ,
\end{equation}
where the three-point tree amplitude is given by
\begin{equation}
 M_3^{(0)}(1^+,2^+,3^-)=-i\left( i\frac{[12]^3}{[23][31]} \right)^2\ .
\end{equation}
Substituting this tree-level amplitude and the one-loop result (\ref{eq:4ptallplus})
into (\ref{eq:5ptallplus1}) yields
\begin{equation}
-\frac{i}{(4 \pi)^2} \frac{1}{60}
\frac{[34]^4[12]^5}{\lan12\ran} \frac{\big(\lan\hat3 4\ran^2 [34]^2+\lan\hat3 4\ran[34]\lan45\ran[45]+\lan45\ran^2[45]^2\big)}
{\lan5|\hat{K}|1]^2\lan5|\hat{K}|2]^2}\ . \nonumber
\end{equation}
We can eliminate $\hat{K}_{12}$ from this expression using
$\lan5|\hat{K}|1]^2=\lan\hat2 5\ran^2 [12]^2$ and $\lan5|
\hat{K}|2]^2=\lan\hat1 5\ran^2 [12]^2$.
Figure \ref{figure-5ptallplus1} gives $\lan\hat{1}\hat{2}\ran\!=\!0$ which corresponds to a pole
 in the complex $z$-plane at
$z=- \lan12\ran / \lan\eta|1+2|3] $.
Then,
using $\eta = \lambda_4 +\lambda_5$ gives the final contribution
 from Figure \ref{figure-5ptallplus1} and (\ref{eq:5ptallplus1}),
\begin{eqnarray}
&&-\frac{i}{(4 \pi)^2} \frac{1}{60}
\frac{[12]\big([34]-[35]\big)^4}{\lan12\ran \big(\lan14\ran+\lan15\ran\big)^2 \big(\lan24\ran+\lan25\ran\big)^2}
\Bigg\{ \left( \lan34\ran -\frac{\lan12\ran[12]}{[34]-[35]} \right)^2[34]^2 \nonumber \\
&& \hskip4.5cm + \left( \lan34\ran -\frac{\lan12\ran[12]}{[34]-[35]} \right)
[34]\lan45\ran[45] + \lan45\ran^2[45]^2 \Bigg\}\ .
\end{eqnarray}
The contributions from the diagrams corresponding to the poles $\lan\hat{2}\hat{3}\ran\!=\!0$
and $\lan\hat{3}\hat{1}\ran\!=\!0$ are then given by cyclically permuting the external legs $\{1,2,3\}$.

We now consider diagrams of the second type.
The diagram in  Figure \ref{figure-5ptallplus2} contributes
\begin{equation}
-\frac{i}{(4 \pi)^2} \frac{1}{60}
\frac{[23]^4[15]^5}{\lan15\ran} \frac{\big(\lan\hat2\hat3\ran^2 [23]^2+\lan\hat2\hat3\ran[23]\lan\hat34\ran[34]+
\lan\hat34\ran^2[34]^2\big)}{\lan4|\hat{K}|1]^2\lan4|\hat{K}|5]^2}\ . \nonumber
\end{equation}
Then, using $\eta = \lambda_4 +\lambda_5$ gives the contribution from Figure
 \ref{figure-5ptallplus2},
\begin{eqnarray}
&&-\frac{i}{(4 \pi)^2} \frac{1}{60}
\frac{[15][23]^4}{\lan15\ran \lan45\ran^2 \big(\lan14\ran+\lan15\ran\big)^2}\Bigg\{ \Big(\lan23\ran[23]+\lan15\ran ([14]-[15])\Big)^2 \nonumber \\
&& \hskip4cm + \Big(\lan23\ran[23]+\lan15\ran ([14]-[15])\Big) \left(\lan34\ran[34]+\frac{\lan15\ran[12][34]}{[23]} \right) \nonumber \\
&& \hskip4cm + \left(\lan34\ran[34]+ \frac{\lan15\ran[12][34]}{[23]} \right)^2 \Bigg\}\ .
\end{eqnarray}
The contributions from the diagrams corresponding to the poles $\lan\hat25\ran\!=\!0$ and
$\lan\hat35\ran\!=\!0$ are obtained by cyclically permuting the external legs $\{1,2,3\}$.
The diagram corresponding to the pole $\lan\hat14\ran\!=\!0$ is obtained from the
$\lan\hat15\ran\!=\!0$ diagram by interchanging legs $4$ and $5$. The remaining diagrams
corresponding to the poles $\lan\hat24\ran\!=\!0$ and $\lan\hat34\ran\!=\!0$ are
then obtained by cyclically permuting the external legs $\{1,2,3\}$ of the
$\lan\hat14\ran\!=\!0$ diagram.

We have checked numerically that each of the terms in the recursion relation
agree with the residues of the expression obtained by shifting the known answer (\ref{eq:allplus})
using the shifts \eqref{eq:risagershifts}. Hence
the sum of the nine recursion relation terms is in precise agreement with the thirty terms
of the known answer (\ref{eq:allplus}).

\subsection{The six-point all-plus amplitude}

\begin{figure}[ht]
\begin{center}
\scalebox{0.65}{\includegraphics{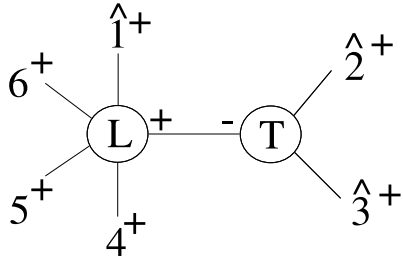}}
\end{center}
\caption{\it
The diagram in the recursive expression for $M_5^{(1)}(1^+,2^+,3^+,4^+,5^+,6^+)$ associated with the pole $\lan\hat{2}\hat{3}\ran=0$.}
\label{figure-6ptallplus1}
\end{figure}

Next we consider the six-point all-plus amplitude.
We again use the  shifts (\ref{eq:risagershifts}), but now choose
$\eta=\lambda_4+\lambda_5+\lambda_6$. Just as in the previous five-point case, all diagrams
contain a one-loop all-plus amplitude and a $++-$ tree-level amplitude, and there are again
two types of diagrams. The first type corresponds to having two shifted legs attached
to the three-point tree-level amplitude. There are three such diagrams, associated with the
three poles $\lan\hat1\hat2\ran\!=\!\lan\hat2\hat3\ran\!=\!\lan\hat3\hat1\ran\!=\!0$.
The diagram associated with the pole $\lan\hat2\hat3\ran\!=\!0$ is given in Figure
\ref{figure-6ptallplus1}.

\begin{figure}[ht]
\begin{center}
\scalebox{0.65}{\includegraphics{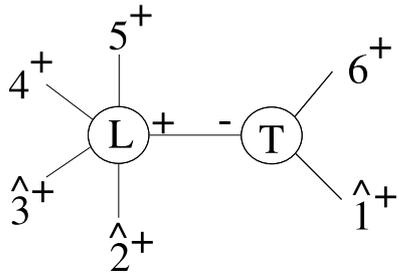}}
\end{center}
\caption{\it
The diagram in the recursive expression for $M_5^{(1)}(1^+,2^+,3^+,4^+,5^+,6^+)$ associated with the pole $\lan\hat{1}6\ran=0$.}
\label{figure-6ptallplus2}
\end{figure}

The second type of diagram corresponds to having a shifted and an unshifted leg attached to the
three-point tree-level amplitude. There are nine such diagrams,  associated with
the nine poles
$\lan\hat14\ran\!=\!\lan\hat15\ran\!=\!\lan\hat16\ran\!=\!\lan\hat24\ran\!=
\!\lan\hat25\ran\!=\!\lan\hat26\ran\!=\!\lan\hat34\ran\!=\!\lan\hat35\ran\!=\!\lan\hat36\ran\!=\!0$.
The diagram associated with the pole $\lan\hat16\ran\!=\!0$ is
given in Figure \ref{figure-6ptallplus2}.

Let us start by calculating the recursive diagram in Figure \ref{figure-6ptallplus1}.
This contributes
\begin{equation}\label{eq:6ptallplus1}
M_5^{(1)}(4^+,5^+,6^+,\hat{1}^+,\hat{K}_{23}^+) \frac{i}{K_{23}} M_3^{(0)}
(\hat{2}^+,\hat{3}^+,-\hat{K}_{23}^-)\ .
\end{equation}
The $M_5^{(1)}(+++++)$ amplitude contains thirty terms (\ref{eq:allplus}). Fortunately the symmetries
in the shifts and the choice for $|\eta\ran\!=\!|4\ran+|5\ran+|6\ran$
reduce these thirty terms to the following ten terms (plus the two cyclic
permutations involving $\{4,5,6\}$ of these ten terms):
\begin{eqnarray}
&&\frac{8i}{(4\pi)^2} \left[ -\frac{[23] \lan46\ran [46]^3 \lan56\ran [56]^3 \big(\lan45\ran[45]+\lan56\ran[56]+\lan46\ran[46]\big)}{\lan23\ran \lan\hat14\ran \lan\hat1 5\ran \lan\hat24\ran \lan\hat25\ran 
\lan \hat{1} \hat{3}\ran^2
} \right. \nonumber\\
&&\phantom{\frac{8i}{(4\pi)^2}}- \frac{[23][14]^3[15]^3 \lan\hat14\ran \lan\hat15\ran \big(\lan\hat14\ran[14]+\lan\hat15\ran[15]+\lan45\ran[45]\big)}{\lan23\ran \lan46\ran \lan56\ran \lan\hat24\ran \lan\hat25\ran \lan\hat36\ran^2} \nonumber\\
&&\phantom{\frac{8i}{(4\pi)^2}}- \frac{[23][16] \big(\lan\hat24\ran[24]+\lan\hat34\ran[34]\big)^3 \big(\lan\hat25\ran[25]+\lan\hat35\ran[35]\big)^3}{\lan23\ran \lan46\ran \lan56\ran \lan\hat14\ran \lan\hat15\ran \lan\hat16\ran \lan\hat34\ran^2 \lan\hat25\ran^2} \nonumber\\
&&\phantom{\frac{8i}{(4\pi)^2}}+ \frac{[23] \lan45\ran [45]^3 [15]^3 \lan\hat15\ran \big(\lan\hat14\ran[14]+\lan\hat15\ran[15]+\lan45\ran[45]\big)}{\lan23\ran \lan46\ran \lan\hat16\ran \lan\hat24\ran \lan\hat1\hat2\ran \lan\hat36\ran^2} \nonumber \\
%
&&\phantom{\frac{8i}{(4\pi)^2}} + \frac{[23] \lan46\ran [46]^3 [16]^3 \lan\hat16\ran \big(\lan\hat14\ran[14]+\lan\hat16\ran[16]+\lan46\ran[46]\big)}{\lan23\ran \lan45\ran \lan\hat15\ran \lan\hat24\ran \lan\hat1\hat2\ran \lan\hat35\ran^2}\nonumber\\
&&\phantom{\frac{8i}{(4\pi)^2}} - \frac{[23][56] \big(\lan45\ran[45]+\lan56\ran[56]+\lan46\ran[46]\big)^3 \big(\lan\hat24\ran[24]+\lan\hat34\ran[34]\big)^3}{\lan23\ran \lan45\ran \lan46\ran \lan56\ran \lan\hat15\ran \lan\hat16\ran \lan\hat3\hat1\ran^2 \lan\hat24\ran^2} \nonumber\\
&&\phantom{\frac{8i}{(4\pi)^2}} - \frac{[23][16]\lan45\ran[45]^3 \big(\lan\hat25\ran[25]+\lan\hat35\ran[35]\big)^3}{\lan23\ran \lan46\ran \lan\hat14\ran \lan\hat16\ran \lan\hat26\ran \lan\hat1\hat2\ran \lan\hat35\ran^2} \nonumber\\
&&\phantom{\frac{8i}{(4\pi)^2}} - \frac{[23][15]\lan46\ran[46]^3 \big(\lan\hat26\ran[26]+\lan\hat36\ran[36]\big)^3}{\lan23\ran \lan45\ran \lan\hat14\ran \lan\hat15\ran \lan\hat25\ran \lan\hat1\hat2\ran \lan\hat36\ran^2} \nonumber\\
&&\phantom{\frac{8i}{(4\pi)^2}} - \frac{[23][56][14]^3 \lan\hat14\ran \big(\lan45\ran[45]+\lan56\ran[56] +\lan46\ran[46]\big)^3}{\lan23\ran \lan45\ran \lan46\ran \lan56\ran \lan\hat25\ran \lan\hat26\ran \lan\hat3\hat1\ran^2} \nonumber\\
&&\phantom{\frac{8i}{(4\pi)^2}} \left. - \frac{[23][56][14]^3 \lan\hat14\ran \big(\lan\hat24\ran[24]+\lan\hat34\ran[34]\big)^3}{\lan23\ran \lan56\ran \lan\hat15\ran \lan\hat16\ran \lan\hat25\ran \lan\hat26\ran \lan\hat34\ran^2} \right] \ . \label{eq:6ptallplusresult1}
\end{eqnarray}
The diagram in   figure \ref{figure-6ptallplus1}   is  
associated with $\lan\hat{2}\hat{3}\ran\!=\!0$ or, equivalently,  with
$z=-\lan23\ran / \lan\eta|2+3|1]$.
Using this value for $z$ it is then simple to rewrite the brackets
containing hatted spinors in (\ref{eq:6ptallplusresult1}) in terms of the
unshifted spinor variables. We have checked numerically that the expression
\eqref{eq:6ptallplusresult1} plus the permutations agrees with the residue at
$\lan\hat{2}\hat{3}\ran\!=\!0$ of the known amplitude
\eqref{eq:allplus}.

The prototypical diagram of the second type is drawn
in Figure \ref{figure-6ptallplus2}. This contributes
\begin{equation}\label{eq:6ptallplus2}
M_5^{(1)}(\hat{2}^+,\hat{3}^+,4^+,5^+,\hat{K}_{16}^+) \frac{i}{K_{16}}
M_3^{(0)}(6^+,\hat{1}^+,-\hat{K}_{16}^-)\ .
\end{equation}
Just like the other term, the $M_5^{(1)}(+++++)$ amplitude
contains thirty terms, but the symmetries of the shifts and
the choice of $| \eta \ran = |4\ran + |5\ran +|6\ran$ reduce
these to the  following set of terms (plus the relevant  permutations):
\begin{eqnarray}\label{eq:6ptset1}
&& \frac{8i}{(4\pi)^2} \left[-\frac{[16] [23]^3 [34]^3 \lan\hat2 \hat3\ran \lan\hat3 4\ran (\lan\hat1 5\ran[15]+ \lan56\ran [56])}{\lan16\ran \lan45\ran \lan56\ran^2 \lan\hat2 5\ran \lan\hat1 \hat2\ran \lan\hat1 4\ran} \right. \nonumber \\
&& \phantom{\frac{8i}{(4\pi)^2}} + \frac{[16] [25]^3 [45]^3 \lan45\ran \lan\hat2 5\ran (\lan\hat3 \hat1\ran[31] + \lan\hat3 6\ran [36])}{\lan16\ran \lan\hat2 \hat3\ran \lan\hat3 4\ran \lan\hat1 4\ran \lan\hat1 \hat2\ran \lan\hat3 6\ran^2} \nonumber \\
&& \phantom{\frac{8i}{(4\pi)^2}} + \frac{[16] [35] [24]^3 \lan\hat2 4\ran (\lan\hat1 4\ran [14]+\lan46\ran [46])^3}{\lan16\ran \lan46\ran^2 \lan\hat3 5\ran \lan\hat2 5\ran \lan\hat2 \hat3\ran \lan\hat1 5\ran \lan\hat3 \hat1\ran} \nonumber \\
&& \phantom{\frac{8i}{(4\pi)^2}} - \frac{[16] [35] [24]^3 \lan\hat2 4\ran (\lan\hat1 \hat2\ran[12]+\lan\hat2 6\ran[26])^3}{\lan16\ran \lan45\ran \lan\hat3 4\ran \lan\hat3 5\ran \lan\hat2 6\ran^2 \lan\hat3 \hat1\ran \lan\hat1 5\ran} \nonumber \\
&& \phantom{\frac{8i}{(4\pi)^2}} + \left. \frac{[16] [35] (\lan\hat1 4\ran [14]+
\lan46\ran [46])^3 (\lan\hat1 \hat2\ran [12] + \lan\hat2 6\ran [26])^3}{\lan16\ran \lan45\ran
\lan\hat3 5\ran \lan\hat2 \hat3\ran \lan\hat2 \hat5\ran \lan\hat3 4\ran \lan\hat2 6\ran^2
 \lan\hat1 4\ran^2} \right] \ .
\end{eqnarray}
We also include three other sets of terms which are the same as \eqref{eq:6ptset1}
but with $(2\leftrightarrow3)$, $(4\leftrightarrow 5)$, and with both $(2\leftrightarrow 3)$
and $(4\leftrightarrow 5)$,
\begin{eqnarray}\label{eq:6ptset2}
&&  \frac{8i}{(4\pi)^2} \left[ \frac{[16] [45] [23]^3 \lan\hat2 \hat3\ran (\lan\hat3 \hat1\ran [31] + \lan\hat3 6\ran [36])^3 }{\lan16\ran \lan45\ran \lan\hat2 4\ran \lan\hat2 5\ran \lan\hat1 4\ran \lan\hat1 5\ran \lan\hat3 6\ran^2} \right. \nonumber\\
&& \phantom{\frac{8i}{(4\pi)^2}} \left. -\frac{[16] [24]^3 [25]^3 \lan\hat2 4\ran \lan\hat2 5\ran (\lan\hat3 \hat1\ran [31]+ \lan\hat3 6\ran [36])}{\lan16\ran \lan\hat3 4\ran \lan\hat3 5\ran \lan\hat1 4\ran \lan\hat1 5\ran \lan\hat3 6\ran^2} \right] \ .
\end{eqnarray}
There is another set of terms which are the same as \eqref{eq:6ptset2}
but with $(2\leftrightarrow 3)$,
\begin{eqnarray}\label{eq:6ptset3}
&& \frac{8i}{(4\pi)^2} \left[ \frac{[16] [24]^3 [34]^3 \lan\hat2 4\ran \lan\hat3 4\ran (\lan\hat15\ran [15] + \lan56\ran [56]) }{\lan16\ran \lan56\ran^2 \lan\hat2 5\ran \lan\hat3 5\ran \lan\hat1 \hat2\ran \lan\hat3 \hat1\ran} \right. \nonumber\\
&& \phantom{\frac{8i}{(4\pi)^2}} \left. +\frac{[16] [23] [45]^3 \lan45\ran (\lan\hat1 5\ran[15]+
\lan56\ran[56])^3}{\lan16\ran \lan56\ran^2 \lan\hat2 4\ran \lan\hat3 4\ran \lan\hat2 \hat3\ran \lan\hat1
\hat2\ran \lan\hat3 \hat1\ran} \right]\ .
\end{eqnarray}
Furthermore, one has another set of terms which are the same as \eqref{eq:6ptset2}
but with $(4\leftrightarrow 5)$,
\begin{eqnarray}\label{eq:6ptset4}
&& \frac{8i}{(4\pi)^2} \left[ \frac{[16] [45] (\lan\hat1 \hat2\ran[12]+\lan\hat2 6\ran[26])^3 (\lan\hat3 \hat1\ran[31]+\lan\hat3 6\ran[36])^3 }{\lan16\ran \lan45\ran \lan\hat2 4\ran  \lan\hat2 5\ran  \lan\hat3 4\ran \lan\hat3 5\ran \lan\hat2 6\ran^2 \lan\hat3 \hat1\ran^2  } \right. \nonumber\\
&& \phantom{\frac{8i}{(4\pi)^2}} \left. - \frac{[16][23](\lan\hat1 4\ran [14]
 + \lan46\ran [46])^3(\lan\hat1 5\ran [15] + \lan56\ran [56])^3}{\lan16\ran  \lan46\ran^2 \lan\hat2 4\ran \lan\hat2 5\ran \lan\hat3
 4\ran \lan\hat3 5\ran \lan\hat1 5\ran^2 \lan\hat2 \hat3\ran} \right]\ .
\end{eqnarray}
Summarising, \eqref{eq:6ptset1} plus permutations contributes 20 terms,
namely
\eqref{eq:6ptset2} plus permutations contributes 4 terms,
\eqref{eq:6ptset3} plus permutations contributes 4 terms and
finally \eqref{eq:6ptset4} contributes 2 terms.
Thus we have a contribution from all 30 terms in the
five-point all-plus amplitude.

The diagram in Figure \ref{figure-6ptallplus2} is associated with $\lan\hat{1} 6\ran\!=\!0$,
or equivalently with $z=-\lan16\ran / [23] \lan\eta 6\ran$.
We have checked numerically that the sum of these terms agrees with the residue of the
shifted known amplitude \eqref{eq:allplus}.
Thus the known six point all-plus gravity amplitude is also correctly reproduced using recursive techniques. It seems likely that this approach will work for all of the all-plus amplitudes.

\section{The one-loop $-+++$ gravity amplitude}

We now turn to study the four-graviton one-loop amplitude with one negative helicity graviton.
This case involves the new feature of double poles in the amplitude,
which introduces complications into the recursion relations. It will be helpful to briefly
review the known results for the gauge theory case before discussing gravity.

In \cite{Bern:2005hs} the five-, six- and seven-point one-loop
Yang-Mills amplitudes with a single negative helicity gluon were
derived from on-shell recursion relations.
Unlike the all-plus amplitude of the previous section, these amplitudes
contain a nonstandard factorisation onto  a
three-point all-plus vertex. The usual collinear limits in real Minkowski space
allow us to derive the leading double-pole factorisation, but
are not precise enough to also calculate the single pole underneath.

\begin{figure}[ht]
\begin{center}
\scalebox{0.6}{\includegraphics{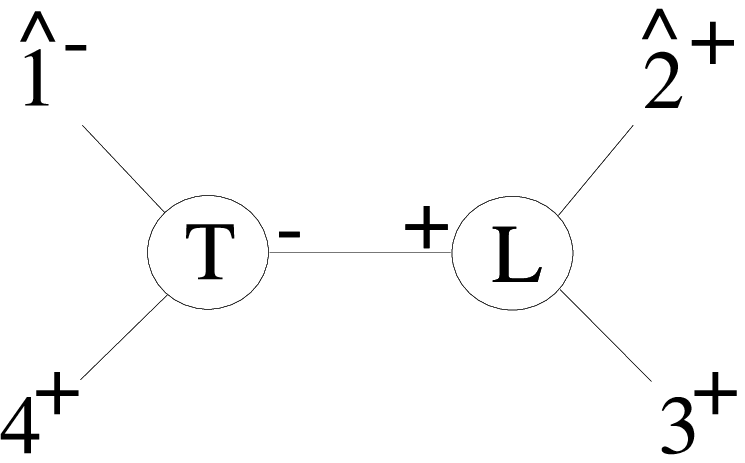}}
\end{center}
\caption{\it The diagram in the recursive expression for $A_4^{(1)}(1^-,2^+,3^+,4^+)$.}\label{figure-4ptympic}
\end{figure}

Consider the simplest four-point case -- the $-+++$ amplitude,
first calculated in \cite{Bern:1991aq},
\begin{equation}\label{eq:4pointym}
A_4^{(1)}(1^-,2^+,3^+,4^+)=\frac{i}{96 \pi^2} \frac{\lan24\ran [24]^3}{[12] \lan23\ran \lan34\ran [41]}\ .
\end{equation}
We will consider recursion based on the standard BCFW shifts on $|1]$ and $|2\ran$:
\begin{eqnarray}
\lambda_1&\to&\lambda_1 \nonumber \\
\widetilde{\lambda}_1&\to&\widetilde{\lambda}_1-z\widetilde{\lambda}_2 \nonumber \\
\lambda_2&\to&\lambda_2+z\lambda_1 \nonumber \\
\widetilde{\lambda}_2&\to&\widetilde{\lambda}_2\ . \label{eq:bcfwshifts1bar2}
\end{eqnarray}
These shifts, when applied to the amplitude, do not give a boundary term
and give a shifted amplitude which is singular at a single point
in the complex $z$-plane,
\begin{eqnarray}
& \lan\hat{2}3\ran = \lan13\ran (z-b) \quad , \quad [\hat{1}4]=  [42] (z-b) \ ,
& \nonumber\\
& {\displaystyle \textrm{where} \quad b = -\frac{\lan23\ran}{\lan13\ran}=\frac{[14]}{[24]}}\ . &
\end{eqnarray}
Applying the shifts (\ref{eq:bcfwshifts1bar2}) to the known amplitude (\ref{eq:4pointym}) yields
\begin{equation}\label{eq:4pointympartialfrac}
A_4^{(1)}(\hat{1}^-,\hat{2}^+,3^+,4^+)(z)=
\frac{i}{96 \pi^2}
\frac{\lan12\ran [24]}{\lan34\ran \lan31\ran}
\left(\frac{1}{(z-b)^2}+\frac{\lan13\ran \lan14\ran}{\lan34\ran \lan12\ran} \frac{1}{(z-b)}\right)\ .
\end{equation}
We can now write the original amplitude $A_4^{(1)}\big(\hat{1}^-,\hat{2}^+,3^+,4^+\big)(0)$
as a sum of residues of the poles that occur in the
function $A_4^{(1)}\big(\hat{1}^-,\hat{2}^+,3^+,4^+\big)(z)/z$.
In this case there is only one contribution, from the residue  at $z=b$.
Following \cite{Bern:2005hs}, this single residue will be explained
recursively by splitting it up into two parts.
The first part comes from the double pole and the
second from the single pole in
(\ref{eq:4pointympartialfrac}).
There will be a one-to-one map between the terms of this expansion and the
terms of a recursion relation based on the shifts (\ref{eq:bcfwshifts1bar2})
\begin{equation}\label{eq:4pointymrecursion}
A_4^{(1)}(1^-,2^+,3^+,4^+)=
\frac{i}{96 \pi^2}
\left[
\frac{\lan12\ran \lan13\ran [24]}{\lan23\ran^2 \lan34\ran}
+\frac{\lan12\ran \lan13\ran [24]}{\lan23\ran^2 \lan34\ran}
\frac{\lan14\ran \lan23\ran}{\lan12\ran \lan34\ran}
\right]\ .
\end{equation}
We recall the origin of these two terms from a recursive diagram --
both are associated with $\lan\hat{2}3\ran=[\hat{1}4]=0$.
The corresponding diagram in presented in Figure \ref{figure-4ptympic}.

BDK \cite{Bern:2005hs} reproduced the double-pole term in (\ref{eq:4pointymrecursion})
recursively from the diagram in  Figure \ref{figure-4ptympic}
using the one-loop all-plus vertex $V_3^{(1)}(+++)$
 in \eqref{eq:ymvertex}
\begin{eqnarray}
A_3^{(0)}(4^+,\hat{1}^+,\hat{K}^-_{23}) \frac{i}{(K_{23}^2)^2} V_{3}^{(1)}(-\hat{K}^+_{23},\hat{2}^+,3^+)
&=&
\frac{i}{96 \pi^2}
\frac{\lan1|\hat{K}|2] \lan1|\hat{K}|3]}{\lan14\ran \lan23\ran^2 [23]}
\frac{\lan\hat{K}1\ran}{\lan\hat{K}4\ran}\  \nonumber \\
&=&
\frac{i}{96 \pi^2}
\frac{\lan12\ran \lan13\ran [24]}{\lan23\ran^2 \lan34\ran}\ .
\end{eqnarray}
The single pole under the double pole term in
(\ref{eq:4pointymrecursion}) differs from the double pole term
by the factor
\begin{equation}
\frac{\lan14\ran \lan23\ran}{\lan12\ran \lan34\ran}\ .
\end{equation}
The BDK correction factor introduced in \eqref{eq:singleunderdouble}
uses the soft functions given in equation \eqref{eq:softfactors}
and in this case is equal to
\begin{equation}
S_3^{(0)}(\hat{1},\hat{K}_{23}^+,4) \, K_{23}^2 \, S_3^{(0)}(\hat{2},-\hat{K}_{23}^-,3)
=\frac{\lan14\ran \lan23\ran [23]^2}{\lan1|\hat{K}|3] \lan4|\hat{K}|2]}
=\frac{\lan14\ran \lan23\ran}{\lan12\ran \lan34\ran}\ .
\end{equation}
We will now see how this approach applies to the gravity case.

The $-+++$ one-loop gravity amplitude which we will re-derive using a recursion relation is
\begin{equation}\label{eq:4ptoneminus}
M_4^{(1)}(1^-,2^+,3^+,4^+)=
\frac{i}{(4\pi)^2}
\frac{1}{180}
\left( \frac{\lan12\ran [23] [24]}{[12] \lan23\ran \lan24\ran} \right)^2
(s^2 + st +t^2)\ ,
\end{equation}
and was calculated using string-based methods in 
\cite{Bern:1993wt}
(we use the normalisation conventions of \cite{Bern:1998sv}).

Remarkably, the recursive procedure for Yang-Mills, reviewed in
the last section, extends very simply to this gravity case.
Just as in the Yang-Mills case,  we consider the standard BCFW shifts on
$|1]$ and $|2\ran$ given in \eqref{eq:bcfwshifts1bar2}.
Applying these shifts to the known amplitude does not give a boundary
term and introduces singularities at two different points in the complex $z$-plane,
\begin{eqnarray}
\lan\hat{2}4\ran&=&\lan14\ran(z-a) \quad\textrm{where}\quad a=-\frac{\lan24\ran}{\lan14\ran}\ ,
\\
\lan\hat{2}3\ran&=&\lan13\ran(z-b) \quad\textrm{where}\quad b=-\frac{\lan23\ran}{\lan13\ran}\ .
\end{eqnarray}
When we reconstruct this amplitude from a recursion
relation the residues at these two points will come from
different diagrams. Of course, there will be  more
recursive diagrams in gravity than there are in Yang-Mills, as
in gravity there is no cyclic ordering of legs like there is for
the colour ordered amplitudes of Yang-Mills.

Under the shifts (\ref{eq:bcfwshifts1bar2}), the amplitude (\ref{eq:4ptoneminus}) becomes
\begin{eqnarray}
M_4^{(1)}\big(\hat{1}^-,\hat{2}^+,3^+,4^+\big)(z)
&=& \frac{i}{(4 \pi)^2} \frac{1}{180} \left[
\frac{\lan12\ran^4[23]^2[24]^2}{\lan13\ran^2 \lan14\ran^2} \frac{1}{(z-a)^2(z-b)^2} 
\right.  \nonumber\\
&&
\phantom{\frac{i}{(4 \pi)^2} \frac{1}{180}} + \frac{\lan12\ran^3[23]^3[24]^2}
{\lan13\ran \lan14\ran^2 [12]} \frac{1}{(z-a)^2(z-b)}\nonumber\\
&&
\phantom{\frac{i}{(4 \pi)^2} \frac{1}{180}}
 + \left. \frac{\lan12\ran^2[23]^4[24]^2}{\lan14\ran^2 [12]^2} \frac{1}{(z-a)^2} \right]
\ .
\end{eqnarray}
We then separate out the different poles using partial fractions.
The shifted amplitude is then expressed as a sum of terms associated
with the various different types of pole at different locations,
\begin{eqnarray}
&&M_4^{(1)}\big(\hat{1}^-,\hat{2}^+,3^+,4^+\big)(z)   \nonumber \\
 && \qquad= \frac{i}{(4 \pi)^2} \frac{1}{180} \bigg[ \frac{\lan12\ran^2 [23]^2 [24]^2}
    {\lan34\ran^2} \frac{1}{(z-a)^2}
      +\frac{\lan12\ran \lan13\ran
   \lan14\ran [23]^2 [24]^2}{\lan43\ran^3} \frac{1}{z-a} \nonumber\\
  &&  \qquad+\frac{\lan12\ran^2 [23]^2
   [24]^2}{\lan34\ran^2} \frac{1}{(z-b)^2}
    + \frac{\lan12\ran
   \lan13\ran \lan14\ran [23]^2
[24]^2}{\lan34\ran^3} \frac{1}{z-b} \bigg] \ .
\end{eqnarray}
Finally, we can write the original amplitude $M_4^{(1)}\big(\hat{1}^-,\hat{2}^+,3^+,4^+\big)(0)$ as a sum of
residues of the  function $M_4^{(1)}\big(\hat{1}^-,\hat{2}^+,3^+,4^+\big)(z)/z$ at the poles of various
types and locations in the complex $z$-plane:
\begin{eqnarray}
M_4^{(1)}&=& \frac{i}{(4 \pi)^2} \frac{1}{180} \left[ \frac{\lan12\ran^2 \lan14\ran^2 [23]^2 [24]^2}{\lan24\ran^2 \lan34\ran^2} \right. \hspace{2.7cm}  \textrm{double-pole, $z=a$} \label{eq:pdresata} \\
&& \phantom{\frac{i}{(4 \pi)^2} \frac{1}{180}} + \frac{\lan12\ran^2 \lan14\ran^2 [23]^2 [24]^2}{\lan24\ran^2 \lan34\ran^2} \left(- \frac{\lan13\ran \lan24\ran}{\lan12\ran \lan43\ran} \right) \textrm{single-pole, $z=a$} \label{eq:spresata} \\
&& \phantom{\frac{i}{(4 \pi)^2} \frac{1}{180}} + \frac{\lan12\ran^2 \lan13\ran^2 [23]^2 [24]^2}{\lan23\ran^2 \lan34\ran^2} \hspace{2.7cm} \textrm{double-pole, $z=b$}\label{eq:dpres}\\
&& \phantom{\frac{i}{(4 \pi)^2} \frac{1}{180}} \left. + \frac{\lan12\ran^2
\lan13\ran^2 [23]^2 [24]^2}{\lan23\ran^2 \lan34\ran^2} \left( - \frac{\lan14\ran
\lan23\ran}{\lan12\ran \lan34\ran}\right) \right] \textrm{single-pole, $z=b$}
\label{eq:spres} 
\end{eqnarray}

\begin{figure}[ht]
\begin{center}
\scalebox{0.6}{\includegraphics{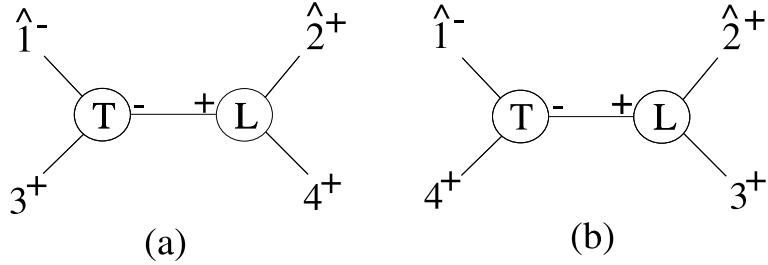}}
\end{center}
\caption{\it
The two diagrams in the recursive expression for $M_4^{(1)}(1^-,2^+,3^+,4^+)$.}\label{figure-4ptpic}
\end{figure}

We will now reconstruct these four terms from the
diagrams of a recursion relation.
There will be two diagrams corresponding to
the two locations, in the complex $z$-plane,
where there are poles in the shifted amplitude
$M_4^{(1)}\big(\hat{1}^-,\hat{2}^+,3^+,4^+\big)(z)$.
The pole at $z\!=\!a$ is associated with
$[\hat{1}3]\!=\!\lan\hat{2}4\ran\!=\!0$
and corresponds to Figure \ref{figure-4ptpic}(a).
The other pole,  at $z\!=\!b$,  is associated with
$[\hat{1}4]\!=\!\lan\hat{2}3\ran\!=\!0$
and corresponds to Figure \ref{figure-4ptpic}(b).

We now introduce a new
three-point one-loop all-plus gravity vertex
\begin{equation}\label{eq:gravityvertex}
W_3^{(1)} (1^+,2^+,3^+)= C ([12][23][31])^2\ ,
\end{equation}
where $C$ is a constant which we will fix shortly by comparison with the known answer (\ref{eq:dpres}).
The double-pole term in (\ref{eq:dpres}) can then be reconstructed
recursively from the diagram corresponding to Figure \ref{figure-4ptpic}b using this.  We have
\begin{equation}\label{eq:dpterm}
M_3^{(0)}(\hat{1}^-,\hat{K}_{23}^{-},4^+)
\frac{i}{(K_{23}^2)^2}
W_3^{(1)}(\hat{2}^+,3^{+},-\hat{K}_{23}^+)
=-C\frac{\lan1|\hat{K}|3]^2 \lan1|\hat{K}|2]^2}{\lan23\ran^2 \lan41\ran^2}
\frac{\lan1\hat{K}\ran^2}{\lan4\hat{K}\ran^2}\ .
\end{equation}

We now use the following relations to write  $\hat{K}$ in terms of the spinor variables of the
external legs,
\begin{eqnarray}
\lan1 | \hat{K} |3]^2&=&\lan1|\hat2 +3|3]^2 = \lan12\ran^2 [23]^2
\ ,
\nonumber\\
\lan1 | \hat{K} |2]^2&=&\lan1|\hat2 +3|2]^2 = \lan13\ran^2 [23]^2
\ ,
\nonumber\\
\frac{\lan1\hat{K}\ran^2}{\lan4\hat{K}\ran^2}&=&\frac{\lan1|\hat{K}|2]^2}{\lan4|\hat{K}|2]^2} = \frac{\lan1|\hat2 +3|2]^2}{\lan4|\hat2 +3|2]^2} = \frac{\lan13
\ran^2}{\lan34\ran^2}
\ .
\end{eqnarray}
Thus (\ref{eq:dpterm}) reproduces the spinor algebra of the known
double pole residue at $z=b$ (\ref{eq:dpres}), giving
\begin{equation}
-C \frac{\lan12\ran^2 \lan13\ran^2 [23]^2 [24]^2}{\lan23\ran^2 \lan34\ran^2}
\ .
\end{equation}
By comparison with (\ref{eq:dpres}) we can fix $C$
in the new vertex $W_3^{(1)}(+++)$ to be:
\begin{equation}\label{eq:valueofC}
C=-\frac{i}{(4\pi)^2}\frac{1}{180}\ .
\end{equation}

The other term (\ref{eq:spres}), corresponding to Figure
\ref{figure-4ptpic}b, is the residue of the
single-pole underneath the double pole at $z=b$.
This single-pole term differs from the double-pole term
(\ref{eq:dpres}) and (\ref{eq:dpterm})
up to a sign in the same way as in the Yang-Mills amplitude
(\ref{eq:4pointymrecursion}):
\begin{equation}
\label{sorpresa}
-\frac{\lan14\ran \lan23\ran}{\lan12\ran \lan34\ran}=
-S_3^{(0)}(\hat{1},\hat{K}_{23}^+,4) K_{23}^2 S_3^{(0)}(\hat{2},-\hat{K}_{23}^-,3)\ .
\end{equation}
Notice that the soft functions in \eqref{sorpresa} are those for Yang-Mills theory (explicitly written in
\eqref{eq:softfactors}), rather than the gravity soft functions.

This remark leads us to suggest the following candidate for the single
pole under the double pole in gravity:
\begin{equation}
-M_3^{(0)}(\hat{1}^-,\hat{K}_{23}^{-},4^+)
S_3^{(0)}(\hat{1},\hat{K}_{23}^+,4)
\frac{i}{K_{23}^2}
S_3^{(0)}(\hat{2},-\hat{K}_{23}^-,3)
W_3^{(1)}(\hat{2}^+,3^{+},-\hat{K}_{23}^+)\ .
\end{equation}
Figure \ref{figure-4ptpic}a is the same as Figure \ref{figure-4ptpic}b,
but with the external legs $3$ and $4$ interchanged.
The two terms (\ref{eq:pdresata}) and (\ref{eq:spresata}), associated
with the residue at  $z\!=\!a$, correspond to Figure \ref{figure-4ptpic}a,
and are similarly found by interchanging legs $3$ and $4$.

\section{The one-loop $-++++$ gravity amplitude}

The $-++++$ one-loop gravity amplitude is unknown.
We now discuss how one might construct it using on-shell recursion relations.
First consider the shifts (\ref{eq:bcfwshifts1bar2}) on $|1]$ and $|2\ran$;
we assume the absence of a boundary term, as
in Yang-Mills, where shifts of the form $[-,+\ran$ have
been observed to be free of large-parameter contributions
\cite{Berger:2006ci,Berger:2006uc}.
Note that this observation extends to the $-+++$ gravity amplitude discussed above.

\begin{figure}[ht]
\begin{center}
\scalebox{0.65}{\includegraphics{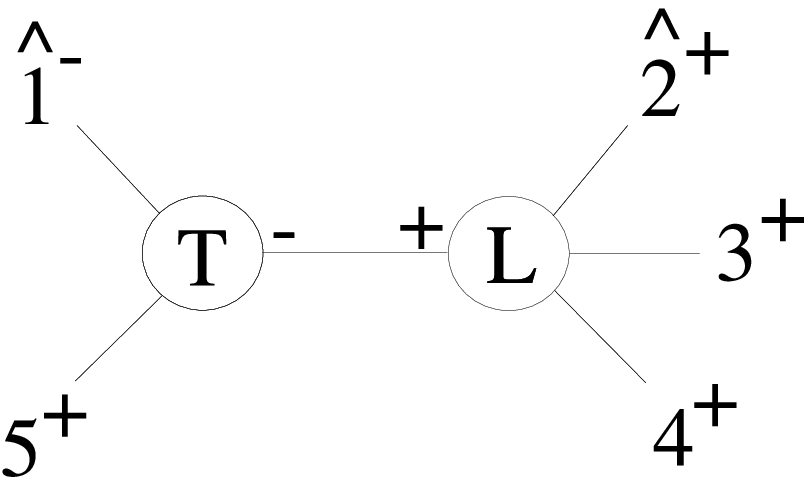}}
\end{center}
\caption{\it The diagram in the recursive expression for $M_5^{(1)}(1^-,2^+,3^+,4^+,5^+)$ corresponding to a simple pole associated with $[\hat1 5]\!=\!0$.}\label{figure-5ptpic1}
\end{figure}

\begin{figure}[ht]
\begin{center}
\scalebox{0.65}{\includegraphics{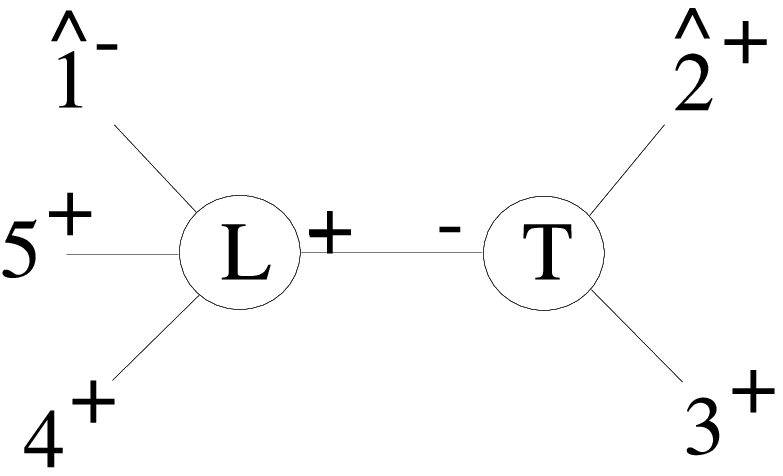}}
\end{center}
\caption{\it The diagram in the recursive expression for $M_5^{(1)}(1^-,2^+,3^+,4^+,5^+)$ corresponding to a simple-pole associated with $\lan\hat23\ran\!=\!0$.}\label{figure-5ptpic2}
\end{figure}

\begin{figure}[ht]
\begin{center}
\scalebox{0.65}{\includegraphics{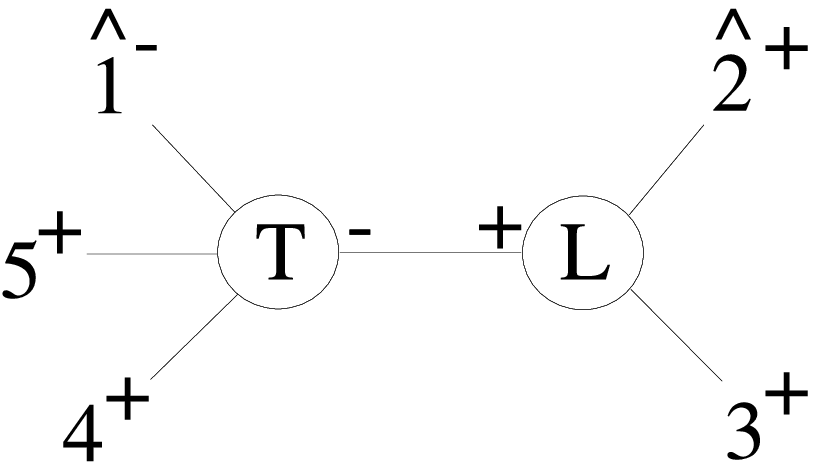}}
\end{center}
\caption{\it The diagram in the recursive expression for $M_5^{(1)}(1^-,2^+,3^+,4^+,5^+)$ corresponding to the double-pole associated with $\lan\hat23\ran\!=\!0$.}\label{figure-5ptpic3}
\end{figure}

The shifts (\ref{eq:bcfwshifts1bar2}) give nine different recursive diagrams.
The shifted amplitude has simple poles associated with
$[\hat13]\!=\![\hat14]\!=\![\hat15]\!=\!0$.
The simple pole associated with $[\hat15]\!=\!0$ corresponds to the
standard recursive diagram in Figure \ref{figure-5ptpic1}.
The shifted amplitude will also have simple poles associated with
$\lan\hat23\ran\!=\!\lan\hat24\ran\!=\!\lan\hat25\ran\!=\!0$.
The simple pole associated with $\lan\hat23\ran\!=\!0$ corresponds
to the standard factorisation diagram in Figure \ref{figure-5ptpic2}.
Finally there are also nonstandard factorisations in the shifted
amplitude corresponding to the poles associated with
$\lan\hat23\ran\!=\!\lan\hat24\ran\!=\!\lan\hat25\ran\!=\!0$.
These nonstandard factorisations contain a one-loop three-point
all-plus vertex,  and contribute double poles and also single poles
under these double poles.
The diagram for the nonstandard factorisation associated
with the pole at $\lan\hat23\ran\!=\!0$ is given in Figure
\ref{figure-5ptpic3}.
There are just three types of diagram to calculate;
Figures \ref{figure-5ptpic1}, \ref{figure-5ptpic2} and \ref{figure-5ptpic3}.
The remaining diagrams can be obtained from these by permuting the
external legs $\{3,4,5\}$.

To begin with, we  consider the diagram in Figure \ref{figure-5ptpic1}.
Using similar manipulations to those detailed in previous sections, one finds that this contributes
\begin{equation}\label{eq:result1}
-\frac{i}{(4\pi)^2}\frac{1}{60}
\frac{\lan15\ran [25]^4}{\lan34\ran^2 [12]^2 [15]}
\Big([23]^2 [45]^2 + [23][34][45][25] + [34]^2[25]^2\Big)\ .
\end{equation}
There is no colour ordering in gravity, so Figure \ref{figure-5ptpic1} is invariant under swapping the external legs labelled $3$ and $4$.
Use of the Schouten identity shows that the result (\ref{eq:result1}) is also invariant under swapping legs $3$ and $4$.

Next we consider Figure \ref{figure-5ptpic2}. This contributes
\begin{equation}\label{eq:result2}
\frac{i}{(4\pi)^2}\frac{1}{180}
\frac{\lan14\ran^2 \lan15\ran^2 [23] [45]^4}{\lan12\ran^2 \lan23\ran  \lan35\ran^2 \lan34\ran^2 \lan45\ran^2}
\Big(\lan14\ran^2 \lan35\ran^2 +\lan14\ran \lan45\ran \lan53\ran \lan13\ran + \lan45\ran^2 \lan13\ran^2\Big)
\, .
\end{equation}
The diagram in Figure \ref{figure-5ptpic2}
is invariant under swapping the external legs labelled $4$ and $5$.
Use of the Schouten identity shows that the result (\ref{eq:result2}) is also invariant under swapping $4$ and $5$.

Finally we consider contributions from the diagram in Figure \ref{figure-5ptpic3}.
This diagram contains a three-point all-plus vertex, hence there will be two
contributions here --
a single and a double pole contribution, as we saw in the four-point example.

First consider the double-pole term,
\begin{equation}\label{eq:diagram3double}
M_4^{(0)} (\hat{1}^-,\hat{K}_{23}^-,4^+,5^+)
\frac{i}{(K_{23}^2)^2}
W_3^{(1)} (\hat{2}^+,3^+,-\hat{K}_{23}^+)\ ,
\end{equation}
where the one-loop three-point all-plus vertex $W_3^{(1)}(+++)$ is the new vertex which was introduced in (\ref{eq:gravityvertex}) and
the $M_4^{(0)}(--++)$ amplitude is given via the following KLT relation
\begin{eqnarray}
M_4^{(0)} (1^-,2^-,3^+,4^+)&=& i \lan12\ran [12] A_4^{(0)}(1^-,2^-,3^+,4^+) A_4^{(0)}(1^-,2^-,4^+,3^+)\nonumber \\
&=& -i\frac{\lan12\ran^7 [12]}{\lan13\ran \lan14\ran \lan23\ran
\lan24\ran \lan34\ran^2}\ .
\end{eqnarray}
Thus (\ref{eq:diagram3double}) yields
\begin{equation}
\label{ll}
C \frac{\lan1|\hat{K}|2]^2 \lan1|\hat{K}|3]^2}{\lan14\ran \lan15\ran \lan23\ran^2 \lan45\ran^2}
\lan1|\hat{K}|\hat{1}] \frac{\lan1\hat{K}\ran \lan1\hat{K}\ran}{\lan4\hat{K}\ran \lan5\hat{K}\ran} \ .
\end{equation}
Eliminating the hats in \eqref{ll}, we obtain
\begin{equation}\label{eq:result3double}
C \frac{\lan12\ran^2 \lan13\ran^4 [23]^4 [45]}{\lan14\ran \lan15\ran
\lan23\ran^2 \lan34\ran \lan35\ran \lan45\ran}\ .
\end{equation}
We recall that the coefficient $C$ has been fixed in \eqref{eq:valueofC}
by comparison with the known $-+++$ one-loop gravity amplitude.
Finally, notice that
Figure \ref{figure-5ptpic3} is invariant under swapping the external legs labelled $4$ and $5$.
The result (\ref{eq:result3double}) is also invariant under swapping $4$ and $5$.

The other contribution from Figure \ref{figure-5ptpic3} is from the
``single-pole underneath the double-pole" term.
Unfortunately this final term poses a problem.
Inspired by the corresponding term (\ref{eq:spres}) in the known $-+++$ gravity
amplitude we might guess that this single-pole under the
double-pole term differs from the double-pole term by a
factor of the form introduced in (\ref{eq:singleunderdouble}), i.e.
%
\begin{equation}
S(a_1,\hat{K}^+,a_2) \, K^2 \, S(b_1,-\hat{K}^-,b_2)\ .
\end{equation}
Experimentation in Yang-Mills \cite{Bern:2005hs} led to the conclusion that
one should choose $a_1$ and $a_2$ to be the legs colour
adjacent to the propagator in the tree-level amplitude of the recursive diagram.
This prescription cannot extend directly to gravity however, since
there is no colour ordering of the external legs.
If we are to use a factor of this form in gravity
we have to choose two of the three external legs attached
to the tree diagram in Figure (\ref{figure-5ptpic3})
to be $a_1$ and $a_2$.

In the $-+++$ gravity amplitude we did not encounter this because
the tree amplitude in the recursive diagram only has two external legs.
However, since a factor of this form is antisymmetric under swapping
$a_1$ and $a_2$, even in the $-+++$ gravity example, the lack of ordering
of the external particles means that this factor has an ambiguous sign.
For Figure (\ref{figure-5ptpic3}) there are three possible choices:
\begin{eqnarray}
S(\hat{1},\hat{K}_{23}^+,4)
K_{23}^2
S(\hat{2},-\hat{K}_{23}^-,3)
&=&-\frac{\lan14\ran \lan23\ran [23]^2}{\lan1|\hat{K}|3] \lan4|\hat{K}|2]}
=\frac{\lan14\ran \lan23\ran}{\lan12\ran \lan34\ran}\ , \\
S(\hat{1},\hat{K}_{23}^+,5)
K_{23}^2
S(\hat{2},-\hat{K}_{23}^-,3)
&=&-\frac{\lan15\ran \lan23\ran [23]^2}{\lan1|\hat{K}|3] \lan5|\hat{K}|2]}
=\frac{\lan15\ran \lan23\ran}{\lan12\ran \lan35\ran}\ , \\
S(5,\hat{K}_{23}^+,4)
K_{23}^2
S(\hat{2},-\hat{K}_{23}^-,3)
&=&\frac{\lan23\ran \lan45\ran [23]^2}{\lan5|\hat{K}|2] \lan4|\hat{K}|3]}
=-\frac{\lan13\ran \lan23\ran \lan45\ran}{\lan12\ran \lan34\ran \lan35\ran}\ .
\end{eqnarray}
It is perhaps natural to guess that a sum of these terms
might give the correct  single pole under the double pole term.
Figure (\ref{figure-5ptpic3}) is symmetric under swapping legs
$4$ and $5$, so we require a sum of factors which share this symmetry.
An appropriate sum of factors would be proportional to
\begin{equation}
\left( \frac{\lan14\ran \lan23\ran}{\lan12\ran \lan34\ran}+
\frac{\lan15\ran \lan23\ran}{\lan12\ran \lan35\ran} \right) \ .
\end{equation}
Collecting together all the previous expressions, and including this term, we are led
to the following proposal for this amplitude ($a$ is a constant)
\begin{eqnarray}
&&M_5^{(1)}(1^-,2^+,3^+,4^+,5^+)\nonumber\\
&&=\frac{i}{(4\pi)^2}\frac{1}{180}\Bigg(
-3
\frac{\lan15\ran [25]^4}{\lan34\ran^2 [12]^2 [15]}
\Big([23]^2 [45]^2 + [23][34][45][25] + [34]^2[25]^2\Big) \nonumber\\
&&
\phantom{\frac{i}{(4\pi)^2}\frac{1}{180}}
+\frac{\lan14\ran^2 \lan15\ran^2 [23] [45]^4}{\lan12\ran^2 \lan23\ran  \lan35\ran^2 \lan34\ran^2 \lan45\ran^2}
\Big(\lan34\ran^2 \lan15\ran^2 +\lan13\ran \lan34\ran \lan45\ran \lan15\ran + \lan13\ran^2 \lan45\ran^2\Big)\nonumber\\
&&\phantom{\frac{i}{(4\pi)^2}\frac{1}{180}}
+\frac{\lan12\ran^2 \lan13\ran^4 [23]^4 [45]}{\lan14\ran \lan15\ran \lan23\ran^2 \lan34\ran \lan35\ran \lan45\ran}
\Big(1+a\frac{\lan14\ran \lan23\ran}{\lan12\ran \lan34\ran}
+
a\frac{\lan15\ran \lan23\ran}{\lan12\ran \lan35\ran} 
\Big)\Bigg)\nonumber\\
&&\nonumber\\
&&
+\frac{i}{(4\pi)^2}\frac{1}{180}\Bigg(
-3
\frac{\lan13\ran [23]^4}{\lan45\ran^2 [12]^2 [13]}
\Big([24]^2 [53]^2 + [24][45][53][23] + [45]^2[23]^2\Big) \nonumber\\
&&
\phantom{\frac{i}{(4\pi)^2}\frac{1}{180}}
+\frac{\lan15\ran^2 \lan13\ran^2 [24] [53]^4}{\lan12\ran^2 \lan24\ran  \lan43\ran^2 \lan45\ran^2 \lan53\ran^2}
\Big(\lan45\ran^2 \lan13\ran^2 +\lan14\ran \lan45\ran \lan53\ran \lan13\ran + \lan14\ran^2 \lan53\ran^2\Big)\nonumber\\
&&
\phantom{\frac{i}{(4\pi)^2}\frac{1}{180}}
+\frac{\lan12\ran^2 \lan14\ran^4 [24]^4 [53]}{\lan15\ran \lan13\ran \lan24\ran^2 \lan45\ran \lan43\ran \lan53\ran}
\Big(1+a\frac{\lan15\ran \lan24\ran}{\lan12\ran \lan45\ran}
+a
\frac{\lan13\ran \lan24\ran}{\lan12\ran \lan43\ran}
\Big)\Bigg)\nonumber\\
&&\nonumber\\
&&+\frac{i}{(4\pi)^2}\frac{1}{180}\Bigg(
-3
\frac{\lan14\ran [24]^4}{\lan53\ran^2 [12]^2 [14]}
\Big([25]^2 [34]^2 + [25][53][34][24] + [53]^2[24]^2\Big) \nonumber\\
&&
\phantom{\frac{i}{(4\pi)^2}\frac{1}{180}}
+\frac{\lan13\ran^2 \lan14\ran^2 [25] [34]^4}{\lan12\ran^2 \lan25\ran  \lan54\ran^2 \lan53\ran^2 \lan34\ran^2}
\Big(\lan53\ran^2 \lan14\ran^2 +\lan15\ran \lan53\ran \lan34\ran \lan14\ran + \lan15\ran^2 \lan34\ran^2\Big)\nonumber\\
&&
\phantom{\frac{i}{(4\pi)^2}\frac{1}{180}}
+\frac{\lan12\ran^2 \lan15\ran^4 [25]^4 [34]}{\lan13\ran \lan14\ran \lan25\ran^2 \lan53\ran \lan54\ran \lan34\ran}
\Big(1+a\frac{\lan13\ran \lan25\ran}{\lan12\ran \lan53\ran}
+a
\frac{\lan14\ran \lan25\ran}{\lan12\ran \lan54\ran}\Big)\Bigg)
\nonumber\\
\label{eq:proposal}
\end{eqnarray}
However, one can check that  this amplitude is not symmetric under the interchange
of legs $2$ and $3$, for any values of the constant $a$,  and hence cannot be the correct answer
as it stands.
We have also checked  that it does not have
all the correct collinear and soft limits for arbitrary  $a$.
Specifically, the (12), and (23), (24), (25) collinear limits are correct, as well as
those (13), (14), (15) collinear limits involving the splitting functions
${\rm Split}^{\rm gravity\, tree }_{-}(i^{+} j^{-})$. However, other limits do not yield the correct results.
Thus one concludes that the methods  reviewed above  fail to work in this case.

Recent papers  may shed light on this problem,  and suggest new approaches
to solve it. The first is the general study on non-standard factorisations of
\cite{Berger:2006ci} (any recursive diagram containing a three-point one-loop part is termed a \lq\lq nonstandard\rq\rq\  factorisation).
It is these factorisations which are the complicating feature in the extension of
the BCFW recursion relation from tree-level amplitudes to the rational parts
of one-loop amplitudes.
For example, factorisations involving the three-point all-plus amplitude give
two types of term, a double pole and a single pole under the double
pole term. While the description of the double pole in terms of a three-point
all-plus vertex appears to be independent of the choice of shifts, the
description of the single pole under the double pole in terms of a multiplicative
correction factor  $S P^2 S$ is not universal even in the Yang-Mills case, as we have checked
in several cases.  It only seems to work for the simplest BCFW shift on $|1]$ and $|2\ran$.

\begin{figure}[ht]
\begin{center}
\scalebox{0.65}{\includegraphics{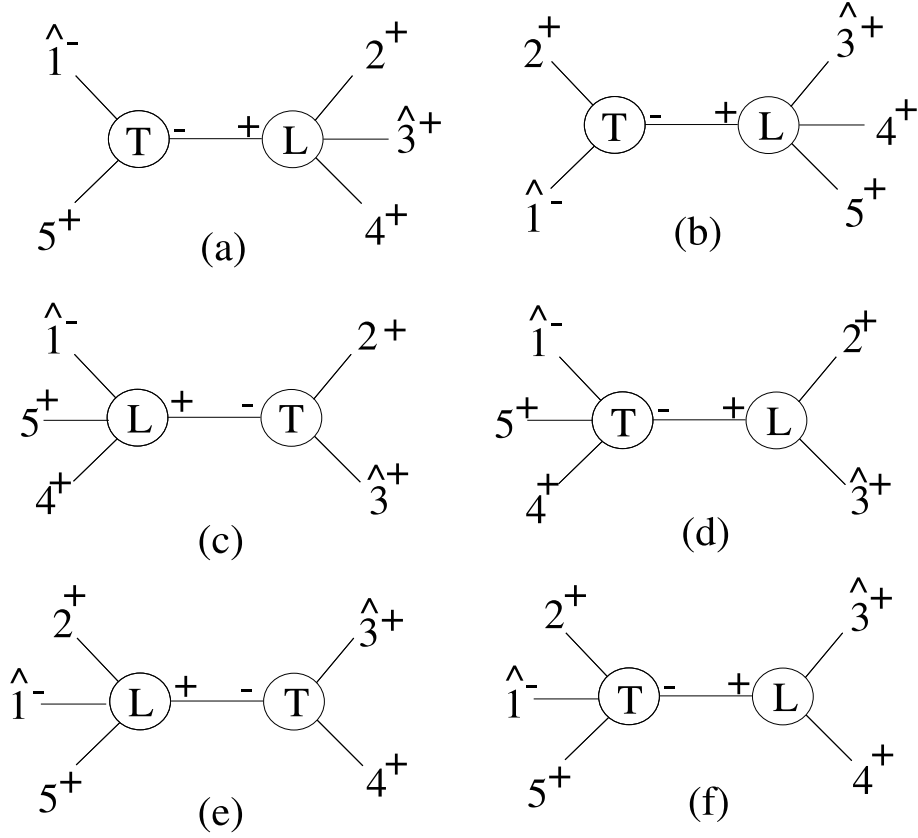}}
\end{center}
\caption{\it The diagrams in the recursive expression for $A_5^{(1)}(1^-,2^+,3^+,4^+,5^+)$.}\label{figure-yangmills}
\end{figure}

We would like to illustrate this point by studying   a specific example in Yang-Mills.
We consider  the $-++++$ Yang-Mills amplitude,  which is given by
\begin{equation}
A_5^{(1)} (1^-,2^+,3^+,4^+,5^+)=i\frac{N_p}{96 \pi^2}
\frac{1}{\lan34\ran^2}
\left[
-\frac{[25]^3}{[12] [51]}
+\frac{\lan14\ran^3 [45] \lan35\ran}{\lan12\ran \lan23\ran \lan45\ran^2}
-\frac{\lan13\ran^3 [32] \lan42\ran}{\lan15\ran \lan54\ran \lan23\ran^2}
\right]\ ,  \label{eq:YMfive}
\end{equation}
and we perform standard BCFW shifts on $|1]$ and $|3\ran$.
We then use partial fractions in order
to separate the various poles. If we then set $z\!=\!0$,  we have
rewritten the amplitude in a form where there is a one-to-one
correspondence between terms in this expansion and the terms of
the recursion relation associated with the shifts:
\begin{eqnarray}
A&=&i\frac{N_p}{96 \pi^2}
\left[ \frac{[35]^3}{\lan24\ran^2 [15] [13]} \right.\label{eq:diaga}\\
&&\hspace{0.5in}+ \frac{[23]^3}{\lan45\ran^2 [12] [13]} \label{eq:diagb}\\
&&\hspace{0.5in}+\frac{\lan25\ran \lan14\ran^3 [45]}{\lan13\ran \lan23\ran \lan45\ran^2 \lan24\ran^2}
\label{eq:diagc}\\
&&\hspace{0.5in}-\frac{\lan12\ran^2 \lan13\ran [23]}{\lan45\ran \lan51\ran \lan24\ran \lan23\ran^2}
\left(1 + 2\frac{\lan14\ran \lan23\ran}{\lan13\ran \lan24\ran}\right) \label{eq:diagd}\\
&&\hspace{0.5in} -\frac{\lan25\ran \lan14\ran^3 [25]}{\lan13\ran \lan34\ran \lan24\ran^2 \lan45\ran^2} \label{eq:diage}\\
&&\hspace{0.5in} \left.-\frac{\lan13\ran \lan14\ran^3 [43]}{\lan12\ran \lan51\ran \lan24\ran \lan54\ran \lan34\ran^2}
\left(1-2\frac{\lan12\ran \lan34\ran}{\lan13\ran \lan42\ran}+\frac{\lan15\ran \lan34\ran}{\lan13\ran \lan45\ran}\right) \right] .
\label{eq:diagf}
\end{eqnarray}
The diagrams in one-to-one correspondence with the above terms
are given in Figure \ref{figure-yangmills}.
The terms (\ref{eq:diaga})--(\ref{eq:diagf})   correspond to diagrams
 \ref{figure-yangmills}(a)--\ref{figure-yangmills}(f) respectively.

We find that the recursive description of these terms is well understood with the
exception of the two factors relating the single pole under double pole
terms to the corresponding double pole terms. Specifically, we require explanations of
the factor  $2 {\lan14\ran \lan23\ran}/({\lan13\ran \lan24\ran})$
in (\ref{eq:diagd}) for the single pole under
the double pole at $\lan2\hat3\ran\!=\!0$, and the
factor
$-2 {\lan12\ran \lan34\ran}/( {\lan13\ran \lan42\ran})+{\lan15\ran \lan34\ran}/ ({\lan13\ran \lan45\ran})$
 in (\ref{eq:diagf}) for the single pole under the double pole at
$\lan\hat34\ran\!=\!0$.

The spinor algebra in the single pole under double pole factor in
\eqref{eq:diagd} might be explained in a style similar to  \cite{Bern:2005hs} by looking at
the diagram in Figure \ref{figure-yangmills}(d) and considering
the legs colour adjacent to the propagator:
\begin{equation}
S(\hat1,k^+,4) K_{23} S(2,k^-,\hat3)=
 \frac{\lan14\ran}{\lan1\hat{k}\ran \lan\hat{k}4\ran} \lan23\ran [23] \frac{[23]}{[2\hat{k}] [\hat{k}3]}
=  \frac{\lan14\ran \lan23\ran}{\lan13\ran \lan24\ran}\ ,
\end{equation}
although the origin of the factor of 2 that appears in (\ref{eq:diagd}) is not clear.

The spinor algebra that appears in factors in (\ref{eq:diagf}) might similarly
be explained, but for this diagram (Figure \ref{figure-yangmills}(f))
we do not consider the colour adjacent legs to the propagator.
The first factor in (\ref{eq:diagf}) is derived from
\begin{equation}
S(\hat1,k^+,2) K_{34} S(\hat3,k^-,4)=
 \frac{\lan12\ran}{\lan1\hat{k}\ran \lan\hat{k}2\ran} \lan34\ran [34]
\frac{[34]}{[3\hat{k}] [\hat{k}4]}
=\frac{\lan12\ran \lan34\ran}{\lan13\ran \lan24\ran}\ ,
\end{equation}
and the second factor in (\ref{eq:diagf}) is derived from
\begin{equation}
S(\hat1,k^+,5) K_{34} S(\hat3,k^-,4)=
 \frac{\lan15\ran}{\lan1\hat{k}\ran \lan\hat{k}2\ran} \lan34\ran [34]
\frac{[34]}{[3\hat{k}] [\hat{k}4]}
=  \frac{\lan15\ran \lan34\ran}{\lan13\ran \lan45\ran}\ .
\end{equation}

The other possibility is to completely avoid non-standard factorisations.
In Yang-Mills it is possible to find shifts which do not involve non-standard factorisations (although
these will generically lead to a boundary term)
\cite{Berger:2006ci,Berger:2006uc }.
Using a pair of shifts in two independent complex parameters,
these authors exploited this fact to calculate complete amplitudes
avoiding the consideration of any nonstandard factorisations.

We now briefly review their method for the simple
case of a purely rational amplitude.
The pair of shifts are called the primary shift and the auxiliary shift:
\begin{eqnarray}
\textrm{primary shift: } [j,l\ran && \left\{ \begin{array}{l}
\widetilde{\lambda}_j \to \widetilde{\lambda}_j -z \widetilde{\lambda}_l  \\
\lambda_l \to \lambda_l +z \lambda_j \end{array} \right.\\
\textrm{auxiliary shift: } [a,b\ran && \left\{ \begin{array}{l}
\widetilde{\lambda}_a \to \widetilde{\lambda}_a -w \widetilde{\lambda}_b \\
\lambda_b \to \lambda_b +w \lambda_a \end{array} \right.
\end{eqnarray}
The primary shift is chosen to have no non-standard
factorisations, but it has a boundary term, while
the auxiliary shift is chosen to have no boundary term, but it
includes non-standard factorisations.
These two shifts give rise to two recursion relations for the amplitude,
\begin{eqnarray}
A_n^{(1)}(0) &=& \mathop{\textrm{Inf }}_{[j,l\ran} A_n+R_n^{\textrm{D,recursive $[j,l\ran$}} \label{eq:primaryrecursion} \ , \\
A_n^{(1)}(0) &=& R_n^{\textrm{D,recursive $[a,b\ran$}}+R_n^{\textrm{D,non-standard $[a,b\ran$}} \ . \label{eq:auxilaryrecursion}
\end{eqnarray}
We now apply the primary shift to the recursion relation
for the auxiliary shift (\ref{eq:auxilaryrecursion}) to
extract the large $z$ behaviour of the primary shift:
\begin{equation}\label{eq:infauxilaryrecursion}
\mathop{\textrm{Inf }}_{[j,l\ran} A_n = \mathop{\textrm{Inf }}_{[j,l\ran}
 R_n^{\textrm{D,recursive $[a,b\ran$}}+ \mathop{\textrm{Inf }}_{[j,l\ran} R_n^{\textrm{D,non-standard $[a,b\ran$}}\ ,
\end{equation}
where the Inf operation is defined to be the constant term
in the expansion of the shifted term about $z=\infty$.
We wish to avoid calculating terms involving nonstandard
factorisations so we will assume that the following condition holds:
\begin{equation}\label{eq:bergercondition}
\mathop{\textrm{Inf }}_{[j,l\ran} R_n^{\textrm{D,non-standard $[a,b\ran$}}=0\ .
\end{equation}
Since we do not, in general, know how to calculate the terms involving nonstandard
factorisations, it is difficult to check explicitly
if the condition (\ref{eq:bergercondition}) holds for a given pair of shifts. However, if one
calculates an amplitude assuming that (\ref{eq:bergercondition}) holds
and the resulting amplitude has the correct collinear and soft behaviour,
then the amplitude is likely to be correct,
and the property (\ref{eq:bergercondition})
must then have been true.
Thus, if we assume the condition (\ref{eq:bergercondition})
and use (\ref{eq:infauxilaryrecursion}) to calculate the boundary
term in (\ref{eq:primaryrecursion}) we can calculate the amplitude
without considering any nonstandard factorisations:
\begin{equation}\label{bergerconstruction}
A_n(0)=  \mathop{\textrm{Inf }}_{[j,l\ran} R_n^{\textrm{D,recursive $[a,b\ran$}} + R_n^{\textrm{D,recursive $[j,l\ran$}}\ .
\end{equation}

\begin{figure}[ht]
\begin{center}
\scalebox{0.6}{\includegraphics{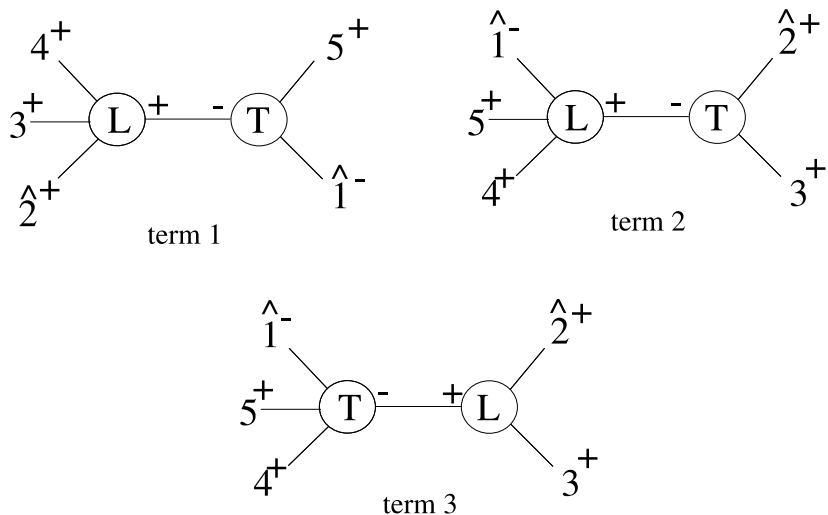}}
\end{center}
\caption{\it The diagrams in the $|1]$ $|2\ran$ shift of $A_5^{(1)}(1^-,2^+,3^+,4^+,5^+)$.}\label{figure-aux5point}
\end{figure}

The following simple example exhibits the possibility of calculating an
amplitude using auxiliary recursions to avoid all nonstandard factorisations
(this is related to examples given in \cite{Berger:2006ci}).
The three terms in the five point Yang-Mills amplitude (\ref{eq:YMfive}) above will be called term 1,
term 2 and term 3 for the
purposes of this section. As shown in \cite{Bern:2005hs}, if we consider
the standard BCFW shifts on $|1]$ and $|2\ran$ then term 1 and term 2
come from standard factorisations and term 3 comes from a nonstandard
factorisation (see Figure \ref{figure-aux5point}). Term 1 comes from the pole associated with $[\hat1 5]=0$, whilst
terms 2 and 3 come from the pole associated with $\lan\hat2 3\ran=0$.
Term 2 is a standard factorisation, but term 3 involves the
nonstandard three-point one-loop all-plus vertex.
In \cite{Bern:2005hs} term 3 was computed by understanding this nonstandard
factorisation as a sum of two terms called a double pole term and
single pole under the double pole term.

Now we show how to use auxiliary recursions to
calculate the amplitude without considering either of the two
types of term associated with the three-point one-loop all-plus nonstandard
factorisations.
We will consider the following pair of shifts.
The primary $[j,l\ran$ shift is on $|4]$ and $|5\ran$.
This shift has no nonstandard factorisations, but does have
boundary term.
The auxiliary $[a,b\ran$ shift is on $|1]$ and $|2\ran$.
This shift has no boundary term, but does have
nonstandard factorisations.
From the discussion in the previous paragraph we known that
\begin{eqnarray}
R_n^{\textrm{D,recursive $[a,b\ran$}}&=& \textrm{term 1 + term 2} \label{eq:thestandardterms} \\
R_n^{\textrm{D,non-standard $[a,b\ran$}} &=& \textrm{term 3} \ .
\end{eqnarray}
Since we are recalculating a known amplitude we can explicitly check
if the condition (\ref{eq:bergercondition}) is satisfied.
If we perform the $[j,l\ran$ shift on term 3 (put hats on $|4]$ and $|5\ran$)
and then consider large $z$, then the term is $O(1/z)$ so the condition
(\ref{eq:bergercondition}) is satisfied:
\begin{equation}
\mathop{\textrm{Inf }}_{[j,l\ran}R_n^{\textrm{D,non-standard $[a,b\ran$}}=\mathop{\textrm{Inf }}_{[j,l\ran} \textrm{term 3}=0\ .
\end{equation}
Thus it will be possible to calculate the $-++++$ Yang-Mills amplitude
without considering any nonstandard factorisations using this
pair of shifts.

\begin{figure}[ht]
\begin{center}
\scalebox{0.6}{\includegraphics{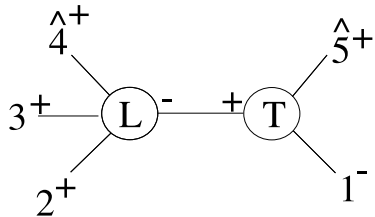}}
\end{center}
\caption{\it The diagram in the $|4]$ $|5\ran$ shift of $A_5^{(1)}(1^-,2^+,3^+,4^+,5^+)$.}\label{figure-pri5point}
\end{figure}

Now we summarise the details of actually calculating the amplitude.
First we use (\ref{eq:thestandardterms}) to calculate the
first term in (\ref{bergerconstruction}).
\begin{equation}\label{eq:firstbit}
\mathop{\textrm{Inf }}_{[j,l\ran} R_n^{\textrm{D,recursive $[a,b\ran$}}=
\mathop{\textrm{Inf }}_{[j,l\ran} \Big(\textrm{term 1 + term 2}\Big)= \textrm{term 1 + term 2}\ .
\end{equation}
As explained earlier, this term should be
thought of as the boundary term in the primary shift.
Finally we have to calculate the recursive diagrams in the
primary $[j,l\ran$ shift on $|4]$ and $|5\ran$.
There is only one diagram associated with these shifts.
This is the diagram corresponding to a pole at $\lan1 \hat5\ran=0$
(see Figure \ref{figure-pri5point}).
Calculating this diagram gives the third term in
(\ref{bergerconstruction})
\begin{equation}\label{eq:secondbit}
R_n^{\textrm{D,recursive $[j,l\ran$}}=\textrm{term 3}\ .
\end{equation}
Thus, putting (\ref{eq:firstbit}) and (\ref{eq:secondbit}) into
the equation (\ref{bergerconstruction}) constructs the full amplitude.
\begin{equation}
A_n(0)=\textrm{term 1}+\textrm{term 2}+\textrm{term 3}
\end{equation}

However, when we try to follow the above example
in order to calculate the one-loop $-++++$ gravity amplitude this proves unsuccessful.
In gravity,  a shift generally involves more factorisations since
there is no cyclic ordering condition on the external
legs.
Using the primary shift on $|4]$ and $|5\ran$ will not work for
gravity since some of these extra factorisations are nonstandard.
The shift on $|4]$ and $|5\ran$ involves the poles associated with
$\lan\hat5 2\ran=0$ and $\lan\hat5 3\ran=0$, and these include contributions
from the three-point one-loop all-plus factorisation.

In conclusion,
we have seen that loop level recursion works for gravity in a number of cases, and provides
relatively simple derivations of amplitudes. It seems likely that this will persist for the
all-plus amplitudes in particular. However, in attempting to apply recursion to more complicated cases,
such as the $-++++$ amplitude discussed in this section, one rapidly runs into difficulties.
The currently known methods falter when confronted with the type of double-pole structures encountered
 here; one can derive straightforwardly some of the terms in the amplitude -- such as those given in
(\ref{eq:proposal}). In this case, these terms have a number of correct properties -- one can check that
some of the collinear and soft limits are correct for example. However, not all limits work, and
neither do the required symmetries (we have also checked that the symmetrisation, in legs $(2,3,4,5)$,
of the expression (\ref{eq:proposal}) does not yield an expression with the right properties). Further
contributions are missing, perhaps involving boundary terms.
It is clear that what is needed is a complete understanding of non-standard
factorisations in complex momenta and a general method of dealing
with double poles.

\section*{Acknowledgements}
It is a pleasure to thank Zvi Bern for helpful
discussions on boundary terms and non-standard factorisations in recursion relations,
and James Bedford and Costas Zoubos for interesting discussions.
We would like to thank PPARC for support under a Rolling Grant PP/D507323/1
and the Special Programme Grant PP/C50426X/1.
The work of GT is supported by an EPSRC Advanced Fellowship and
an EPSRC Standard Research Grant.



\begin{thebibliography}{99}

\bibitem{Witten:2003nn}
E.~Witten,
{\it Perturbative gauge theory as a string theory in twistor space},
Commun.\ Math.\ Phys.\  {\bf 252}, 189 (2004),
{\tt hep-th/0312171}.

\bibitem{Cachazo:2005ga}
F.~Cachazo and P.~Svrcek,
{\it Lectures on twistor strings and perturbative Yang-Mills theory},
PoS {\bf RTN2005}, 004 (2005),
{\tt hep-th/0504194}.

\bibitem{Brandhuber:2006vh}
A.~Brandhuber and G.~Travaglini,
{\it Quantum MHV diagrams},
{\tt hep-th/0609011}.

\bibitem{Britto:2004ap}
R.~Britto, F.~Cachazo and B.~Feng,
{\it New recursion relations for tree amplitudes of gluons},
Nucl.\ Phys.\ B {\bf 715}, 499 (2005),
{\tt hep-th/0412308}.

\bibitem{Britto:2005fq}
R.~Britto, F.~Cachazo, B.~Feng and E.~Witten,
{\it Direct proof of tree-level recursion relation in Yang-Mills theory},
Phys.\ Rev.\ Lett.\  {\bf 94}, 181602 (2005),
{\tt hep-th/0501052}.

\bibitem{BDDK}
Z.~Bern, V.~Del Duca, L.~J.~Dixon and D.~A.~Kosower,
{\it All Non-Maximally-Helicity-Violating One-Loop Seven-Gluon Amplitudes In N =
4 Super-Yang-Mills Theory}, Phys. Rev. {\bf D71} (2005) 045006,
{\tt hep-th/0410224}.

\bibitem{BCF}
R.~Britto, F.~Cachazo and B.~Feng,
{\it Generalized unitarity and one-loop amplitudes in N = 4 super-Yang-Mills},
Nucl. Phys. {\bf B725} (2005) 275-305,
{\tt hep-th/0412103}.

\bibitem{BDK}
Z.~Bern, L.~J.~Dixon and D.~A.~Kosower,
{\it All Next-to-Maximally-Helicity-Violating One-Loop Gluon Amplitudes in N=4
Super-Yang-Mills Theory},  Phys. Rev. {\bf D72} (2005) 045014,
{\tt hep-th/0412210}.

\bibitem{RSV}
R.~Roiban, M.~Spradlin and A.~Volovich,
{\it Dissolving N= 4 loop amplitudes into QCD tree amplitudes},
Phys. Rev. Lett. {\bf 94} (2005) 102002,
{\tt hep-th/0412265}.

\bibitem{Luo:2005rx}
M.~x.~Luo and C.~k.~Wen,
{\it Recursion relations for tree amplitudes in super gauge theories},
JHEP {\bf 0503}, 004 (2005),
{\tt hep-th/0501121}.

\bibitem{Luo:2005my}
M.~x.~Luo and C.~k.~Wen,
{\it Compact formulas for all tree amplitudes of six partons},
Phys.\ Rev.\ D {\bf 71}, 091501 (2005),
{\tt hep-th/0502009}.

\bibitem{Badger:2005zh}
S.~D.~Badger, E.~W.~N.~Glover, V.~V.~Khoze and P.~Svrcek,
{\it Recursion relations for gauge theory amplitudes with massive particles},
JHEP {\bf 0507}, 025 (2005),
{\tt hep-th/0504159}.



\bibitem{Bern:2005hs}
Z.~Bern, L.~J.~Dixon and D.~A.~Kosower,
{\it On-shell recurrence relations for one-loop QCD amplitudes},
Phys.\ Rev.\ D {\bf 71}, 105013 (2005),
{\tt hep-th/0501240}.

\bibitem{Bern:2005ji}
Z.~Bern, L.~J.~Dixon and D.~A.~Kosower,
{\it The last of the finite loop amplitudes in QCD},
Phys.\ Rev.\ D {\bf 72}, 125003 (2005),
{\tt hep-ph/0505055}.

\bibitem{Bern:2005cq}
Z.~Bern, L.~J.~Dixon and D.~A.~Kosower,
{\it Bootstrapping multi-parton loop amplitudes in QCD},
Phys.\ Rev.\ D {\bf 73}, 065013 (2006),
{\tt hep-ph/0507005}.

\bibitem{Berger:2006ci}
C.~F.~Berger, Z.~Bern, L.~J.~Dixon, D.~Forde and D.~A.~Kosower,
{\it Bootstrapping one-loop QCD amplitudes with general helicities},
{\tt hep-ph/0604195}.

\bibitem{Berger:2006vq}
C.~F.~Berger, Z.~Bern, L.~J.~Dixon, D.~Forde and D.~A.~Kosower,
{\it All one-loop maximally helicity violating gluonic amplitudes in QCD},
{\tt hep-ph/0607014}.

\bibitem{Bern:2005hh}
Z.~Bern, N.~E.~J.~Bjerrum-Bohr, D.~C.~Dunbar and H.~Ita,
{\it Recursive calculation of one-loop QCD integral coefficients},
JHEP {\bf 0511}, 027 (2005),
{\tt hep-ph/0507019}.

\bibitem{Bedford:2005yy}
J.~Bedford, A.~Brandhuber, B.~J.~Spence and G.~Travaglini,
{\it A recursion relation for gravity amplitudes},
Nucl.\ Phys.\ B {\bf 721}, 98 (2005),
{\tt hep-th/0502146}.

\bibitem{Cachazo:2005ca}
F.~Cachazo and P.~Svrcek,
{\it Tree level recursion relations in general relativity},
{\tt hep-th/0502160}.

\bibitem{Bjerrum-Bohr:2006yw}
  N.~E.~J.~Bjerrum-Bohr, D.~C.~Dunbar, H.~Ita, W.~B.~Perkins and K.~Risager,
  {\it The no-triangle hypothesis for N = 8 supergravity},
{\tt hep-th/0610043}.

\bibitem{Green:2006gt}
  M.~B.~Green, J.~G.~Russo and P.~Vanhove,
  {\it Non-renormalisation conditions in type II string theory and maximal supergravity},
  {\tt hep-th/0610299}.

\bibitem{Bern:2006kd}
  Z.~Bern, L.~J.~Dixon and R.~Roiban,
  {\it Is N = 8 supergravity ultraviolet finite?},
  {\tt hep-th/0611086}.

\bibitem{Green:2006yu}
  M.~B.~Green, J.~G.~Russo and P.~Vanhove,
  {\it Ultraviolet properties of maximal supergravity,}
  {\tt hep-th/0611273}.


\bibitem{Risager:2005vk}
K.~Risager,
{\it A direct proof of the CSW rules},
JHEP {\bf 0512}, 003 (2005),
{\tt hep-th/0508206}.


\bibitem{csw}
F.~Cachazo, P.~Svrcek and E.~Witten,
{\it MHV vertices and tree amplitudes in gauge theory},
JHEP {\bf 0409}, 006 (2004),
{\tt hep-th/0403047}.




\bibitem{Bern:1998xc}
Z.~Bern, L.~J.~Dixon, M.~Perelstein and J.~S.~Rozowsky,
{\it One-loop n-point helicity amplitudes in (self-dual) gravity},
Phys.\ Lett.\ B {\bf 444}, 273 (1998),
{\tt hep-th/9809160}.

\bibitem{Bern:1998sv}
Z.~Bern, L.~J.~Dixon, M.~Perelstein and J.~S.~Rozowsky,
{\it Multi-leg one-loop gravity amplitudes from gauge theory},
Nucl.\ Phys.\ B {\bf 546}, 423 (1999),
{\tt hep-th/9811140}.



\bibitem{Dunbar:1994bn}
D.~C.~Dunbar and P.~S.~Norridge,
{\it Calculation of graviton scattering amplitudes using string based methods},
Nucl.\ Phys.\ B {\bf 433}, 181 (1995),
{\tt hep-th/9408014}.




\bibitem{Bern:1991aq}
Z.~Bern and D.~A.~Kosower,
{\it The Computation of loop amplitudes in gauge theories},
Nucl.\ Phys.\ B {\bf 379}, 451 (1992).



%
\bibitem{Bern:1993wt}
  Z.~Bern, D.~C.~Dunbar and T.~Shimada,
  {\it String based methods in perturbative gravity,}
  Phys.\ Lett.\ B {\bf 312} (1993) 277, 
  {\tt hep-th/9307001}.

\bibitem{Berger:2006uc}
C.~F.~Berger,
{\it Bootstrapping one-loop QCD amplitudes},
{\tt hep-ph/0608027}.


\end{thebibliography}
\end{document}